\documentclass[useAMS,usenatbib]{mn2e}
\usepackage{rotating}
\usepackage{color}
\usepackage{amsmath}
\usepackage{float}
\usepackage{graphicx}
\usepackage{fixltx2e}
\usepackage{dblfloatfix}

\newcommand{\Msun}{M$_{\odot}$}

\newcommand{\FW}{{\sc{FellWalker}}}

\newcommand{\gsim}{\;\lower.6ex\hbox{$\sim$}\kern-7.75pt\raise.65ex\hbox{$>$}\;}

\title[The JCMT Plane Survey]{The JCMT Plane Survey: early results from the $\ell=30^{\rm o}$ field}

\author[
T. J. T. Moore et al.]{T. J. T. Moore$^{1}$\thanks{E-mail:
T.J.Moore@ljmu.ac.uk (TJTM)} 
R.\ Plume$^{2}$, M. A. Thompson$^{3}$,
 H. Parsons$^{4}$, J. S.\ Urquhart$^{5}$, \and D. J. Eden$^{1,6}$, J. T. Dempsey$^{4}$, L. K. Morgan$^{1}$, 
 H. S. Thomas$^{4}$,  
J. Buckle$^{7,8}$, \and C. M. Brunt$^{9}$, H. Butner$^{10}$, D. Carretero$^{8}$, A. Chrysostomou$^{3}$, H. M. deVilliers$^{3}$, \and M. Fich$^{11}$, M. G. Hoare$^{12}$, G. Manser$^{3}$, J. C. Mottram$^{13}$, C. Natario$^{3}$, F. Olguin$^{12}$, \and N. Peretto$^{14}$, D. Polychroni$^{15}$, R. O. Redman$^{4}$, A. J. Rigby$^{1}$, C. Salji$^{8}$, \and L. J. Summers$^{9}$, 
D. Berry$^{4}$, M. J. Currie$^{4}$, T. Jenness$^{4,16}$, 
M. Pestalozzi$^{17}$, \and A. Traficante$^{18}$, 
P. Bastien$^{19}$,  J. diFrancesco$^{20}$, C. J. Davis$^{1}$, A. Evans$^{21}$,  \and P. Friberg$^{4}$, G. A. Fuller$^{18}$,  A. G. Gibb$^{22}$, S. J. Gibson$^{23}$,  T. Hill$^{24}$,  D. Johnstone$^{4,20,25}$, \and G. Joncas$^{26}$, S. N. Longmore$^{1}$, S. L. Lumsden$^{12}$, P. G. Martin$^{27}$, \and Q. 
Nguy$\tilde{\hat{\rm e}}$n Lu'o'ng$^{27}$
J. E. Pi\~neda$^{18}$, C. Purcell$^{28}$,  J. S. Richer$^{8}$, G. H. Schieven$^{20}$, \and R.\ Shipman$^{29}$, M. Spaans$^{30}$, A. R. Taylor$^{2}$, S. Viti$^{31}$,  B. Weferling$^{32}$,  G. J. White$^{33,34}$, \and M. Zhu$^{35}$ 
\\ \\
author affiliations are listed on the final page.
}

\begin{document}

\date{Accepted 2015 August 6.  Received 2015 August 5; in original form 2014 December 4}

\pagerange{\pageref{firstpage}--\pageref{lastpage}} \pubyear{2015}

\maketitle

\label{firstpage}

\begin{abstract}
We present early results from the JCMT Plane Survey (JPS), which is surveying the northern inner Galactic Plane between longitudes $\ell=7^{\circ}$ and $\ell=63^{\circ}$ in the 850-$\umu$m continuum with SCUBA-2, as part of the James Clerk Maxwell Telescope Legacy Survey programme.  Data from the $\ell = 30^{\circ}$ survey region, which contains the massive star-forming regions W\,43 and G29.96, are analysed after approximately 40\% of the observations had been completed.  The pixel-to-pixel noise is found to be 19\,mJy\,beam$^{-1}$ after a smooth over the beam area, and the projected equivalent noise levels in the final survey are expected to be around 10\,mJy\,beam$^{-1}$.  An initial extraction of compact sources was performed using the \FW\ method resulting in the detection of 1029 sources above a 5-$\sigma$ surface-brightness threshold.  The completeness limits in these data are estimated to be around 0.2\,Jy\,beam$^{-1}$ (peak flux density) and 0.8\,Jy (integrated flux density) and are therefore probably already dominated by source confusion in this relatively crowded section of the survey.  The flux densities of extracted compact sources are consistent with those of matching detections in the ATLASGAL survey. We analyse the virial and evolutionary state of the detected clumps in the W43 star-forming complex and find that they appear younger than the Galactic-plane average.
\end{abstract}


\clearpage
\section{Introduction}

In order to obtain a predictive understanding of star formation that will be useful to all areas of astrophysics, we must identify and quantify the principal mechanisms responsible for regulating the star-formation rate (SFR) and efficiency (SFE) on different spatial scales, and, potentially, those that determine the stellar IMF.  We need to separate the effects of Galaxy-scale mechanisms, in particular the spiral-arm density waves, from local triggering by feedback due to radiation, stellar winds and H{\sc ii} regions.

Until recently, the study of star formation (and especially massive star formation) has been limited by biased and incomplete samples. For example, what controls the SFE, whether the IMF is variable, or even which regions are typical and which are extreme are all unknowns.  However, these questions are now solvable by using large surveys to measure the star-formation efficiency (SFE) and luminosity function (LF) of YSOs.  Using Red MSX Source (RMS) survey data, \citep{lumsden13}, \citet{moore12} have found that around 70\% of the increase in star-formation rate density associated with spiral arms in the inner Galaxy is due to source crowding, rather than any physical effect of the arms on the star-formation process.  The remainder could be due to increases in SFE or in the mean luminosity of massive YSOs (i.e.\ a flattening of the MYSO luminosity function).  However it might also be the result of residual bias in the source samples due to patchy star formation within spiral arms (\citealp{urquhart14a}).

\citet{moore12} also show evidence that the mean molecular cloud mass decreases with Galactocentric radius, implying a significant change in the star-forming environment with Galactic radius.  The RMS survey is also beginning to unravel the evolutionary timescales of massive young stellar objects (MYSOs).  \citet{mottram11} find that the accretion phase  of MYSOs has a duration ranging from $4\times 10^5$ years for objects with luminosities of $10^4$ L$_\odot$ to $\sim 7\times 10^4$ years at  for those with $L = 10^5$ L$_\odot$.  At luminosities above $10^5$ L$_\odot$ the accretion timescales for MYSOs become comparable to the Kelvin-Helmholtz time.  \citet{davies11} suggest that these timescales are most consistent with accretion rates predicted by turbulent core and competitive accretion models (e.g. \citealt{bonnell06}).

Similarly, there has been recent progress on constraining the role of spiral arms in affecting the efficiency of forming dense, star-forming clumps out of the larger molecular clouds (the clump formation efficiency, or CFE).  Using data from the Galactic Ring Survey (GRS: \citealp{jackson06}) and the Bolocam Galactic Plane Survey (\citealp{aguirre}), in the range $\ell = 37.8^\circ$ to 42.5$^\circ$, \citet{eden13} found no difference in the CFE between the arm and interarm regions.  This implies that the increased star formation rate in Galactic spiral arms arises simply from the presence of greater amounts of gas in the arms, and is not due to an increase in the efficiency of forming dense clumps.

The key data required in such investigations are provided by large-scale surveys of the Galactic Plane in tracers of the cool, dense, molecular gas which forms the initial conditions for star formation. Of particular importance are surveys in rotational emission lines from the isotopologues of CO and continuum emission in the far-infrared and sub- millimetre bands from warm and cold dust mixed with the gas. The thermal sub-millimetre continuum emission from cool and cold dust is associated with a variety of early phases of star formation and is reliably optically thin, giving direct measurements of column density if it is assumed that gas and dust are well mixed. Large-area surveys at these wavelengths therefore provide a census of current and incipient star-formation activity. Unbiased, blind, surveys are required to demonstrate both where star formation is present and where it is absent.

A number of major continuum surveys of the Galactic Plane (GP) have been completed or are in progress, which will allow studies such as those above to be made in an unbiased fashion, with more detail, and over Galactic scales.  These surveys hold out the promise of providing data that will facilitate major steps forward in the study of star formation by providing SED and luminosity information for complete samples of star-forming regions and individual young stellar objects, large enough to bin by luminosity, evolutionary stage and environment.   Among these are: Hi-GAL \citep{molinari10, molinari10b} - an open-time Key Program of the {\em Herschel} Space Observatory, which carried out a 5-band photometric imaging survey at 70, 160, 250, 350, and 500 $\umu$m of a $|b| \le 1^\circ$ wide strip of the entire Galactic plane and following the Galactic warp and ATLASGAL (\citealp{schuller09}), the first systematic survey of the inner Galactic plane at 870\,$\umu$m, covering  $280^\circ < \ell < 60^\circ$ and  $|b| < 1.5^\circ$ over most of its range. This survey was carried out using the LABOCA instrument at the APEX telescope \citealp{siringo}.  CORNISH (\citealp{hoare12} - which has used the VLA to map the northern Galactic plane at 5 GHz in the range $\ell = 10^\circ - 60^\circ$ and at latitudes of $|b| < 1^\circ$.  GLIMPSE (\citealp{benjamin}; \citealp{churchwell}) - which surveyed a range of Galactic latitude and longitude at wavelengths  of 3.6, 4.5, 5.8, and 8.0 $\umu$m.  Specifically $\ell = 10^\circ - 65^\circ$ at $|b| < 1^\circ$, 
the inner $20^\circ$ of the Galactic plane with $b <  2^\circ$ in the inner $\ell = 0^\circ - 5^\circ$, and then nine selected strips extended in latitude to $\pm 3^\circ$ ($\pm 4^\circ$ if they were within $2^\circ$ of the Galactic Centre). 

In this paper, we describe early results from the James Clerk Maxwell Telescope (JCMT) Galactic Plane Survey (JPS),
part of the JCMT Legacy Survey programme \citep{chrysostomou10}.
%\footnote{The JCMT Legacy Surveys programme is also described at http://www.jach.hawaii.edu/JCMT/surveys}.  
The results presented are from a region of the survey near Galactic longitudes of $\ell = 30^{\circ}$, which is well enough studied so that consistency checks can be made and where the source density is high and the detection sensitivity is near to the confusion limit and unlikely to be much improved upon in the final product.  JPS was allocated 450 hours in JCMT weather band 3 ($0.08 \le  \tau_{225\,{\rm GHz}} \le 0.12$) to survey the region $7^\circ \le \ell \le 63^\circ$, $| \, b \, | \le 1^\circ$ in the 850-$\umu$m continuum. Observations began in June 2012 and were completed in late 2014.  The JPS Consortium consists of astronomers affiliated at time of joining to one of the three pre-2013 JCMT partner countries: UK, Canada \& the Netherlands.  JPS is part of a coordinated plan for mapping the Galactic Plane in the sub-mm continuum with JCMT, which involves a second JCMT/SCUBA-2 Legacy-survey project, the SCUBA-2 Ambitious Sky Survey (SASSy; 
%\citealp{thompson14}
Thompson et al., in preparation).  SASSy will observe the outer GP region $120^{\rm o}<\ell<240^{\rm o}$; $|\,b\,| \le 2.5^{\rm o}$ at 850\,$\umu$m with sensitivity 30-40 mJy\,beam$^{-1}$ rms.  The SASSy Perseus extension will also map longitudes $60^{\rm o}<\ell<120^{\rm o}$, completing the SCUBA-2 coverage of the GP visible from JCMT to an approximately constant average mass sensitivity.  JPS covers the same area as the ATLASGAL survey (\citealp{schuller09}) and Bolocam Galactic Plane Survey (BGPS; \citealp{aguirre}) but will be several times deeper.

The goal of the JPS is to produce a census of, principally, massive star formation across a large fraction of the Galaxy, complete to a mass limit of order 100\,M$_{\odot}$ (depending on assumed temperature and emissivity values) at a distance of $\sim$20\,kpc.  Aside from the legacy value of the data, the immediate science aims are to investigate the earliest stages of the evolutionary sequence of massive protostars and pre-stellar objects, the variation of star-formation efficiency with location in the Galaxy and the effect of environment and feedback on this and the mass distribution of dense, star-forming clumps.  A public release of JPS data will accompany a future paper describing the complete survey.

\section{Observing Strategy}

JPS uses the wide-field submillimetre-band bolometer camera SCUBA-2 (the Sub-mm Common-User Bolometer Array 2: \citealp{holland13}) in the 850-$\umu$m band at a spatial resolution of 14.5 arcseconds.

\begin{figure*}
\includegraphics[width=\textwidth,trim=0 0 0 0]{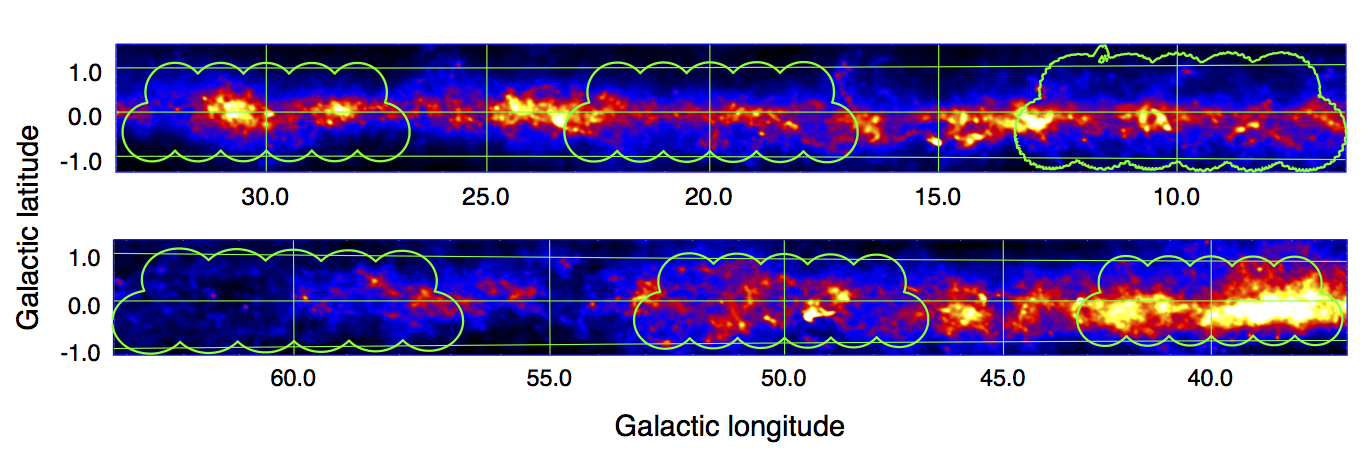}
\caption{JPS projected area coverage on the sky (green contours) overlaid on the Planck dust opacity map (colour scale; \citealp{planck}). The apparent size difference between the six JPS patches is due to the tangent projection of the Galactic coordinate system. The zero of Declination passes through $\ell \simeq 32.9^{\circ}$. The axis scale is degrees.
}
\label{skycover}
\end{figure*}

\begin{figure}
\includegraphics[width=0.47\textwidth,trim=0 0 0 0]{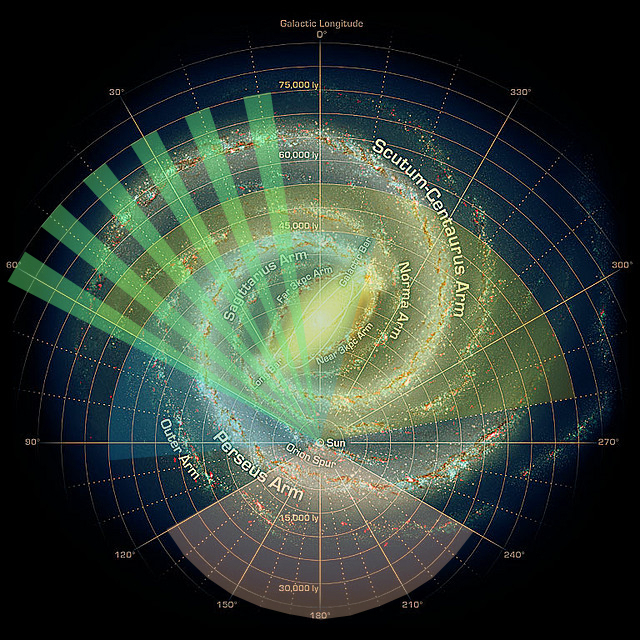}
\caption{The area of the Galaxy covered by JPS (green segments).  Also shown, for comparison, is the longitude coverage of ATLASGAL (yellow), BGPS (blue) and SASSy (pink).  The background image is the sketch of the Milky Way by Robert Hurt of the Spitzer Science Center, made in collaboration with Robert Benjamin.
%{\color{red} add SASSy Per extension and reverse background?}
}
\label{gpplan}
\end{figure}

 JPS will sample the region of the inner GP within the longitude range $7^{\circ}<\ell<63^{\circ}$ and latitude $|\,b\,| \le 0.8^{\circ}$, an area coincident with that of the northern GLIMPSE survey (\citealp{benjamin}, also MIPSGAL: \citealp{carey}), and CORNISH north (\citealp{hoare12, purcell13}).  The strategy is to map six large, regularly spaced fields within this area, centred at intervals of $10^{\circ}$ from $\ell=10^{\circ}$ (Figure \ref{skycover}) with a target rms sensitivity of $\sim$10\,mJy\,beam$^{-1}$ at 850\,$\umu$m.  Each of the fields covers approximately $5^{\circ}$ in longitude and $1.7^{\circ}$ in latitude.  This strategy was developed in response to the re-scoping of the entire JCMT Legacy Survey project that took place in 2011.
The choice of regularly spaced but spatially limited target fields allows a significant increase in depth over existing surveys while preserving the goal of producing a relatively unbiased survey and still taking in several significant features of Galactic structure, particularly the tangent and origin of the Scutum spiral arm at $\ell=30^{\circ}$ and the Sagittarius-arm tangent at $l=50^{\circ}$. 
Figure \ref{gpplan} shows the planned extent of JPS in the context of several other key sub-mm surveys.

The observing modes developed for SCUBA-2 include large-area mapping modes known as rotating {\em pong} patterns in which the telescope tracks across a square area, filling it in by notionally bouncing off the boundary of this area, in the manner of the eponymous vintage video game. Once a pattern is complete the map is rotated and the pattern repeated at the new angle. The combination of telescope velocity, the spacing between successive rows of the basic pattern and the number of rotations have been optimised to give the most uniform exposure-time coverage across the defined field.
%\footnote{more information on SCUBA-2 observing modes can be found at http://www.jach.hawaii.edu/JCMT/continuum/}.  
JPS uses the {\em pong3600} mode \citep{bintley14}, with 8 rotations and a telescope scan speed of 600 arcseconds per second to produce a one-degree circular area of consistent exposure time, to observe each tile in the survey (the actual mapped area is larger than this).  The total time required for each observation is generally between 40 and 45 minutes. 

\begin{figure}
\includegraphics[angle=0,width=0.5\textwidth,trim=0 0 30 0]{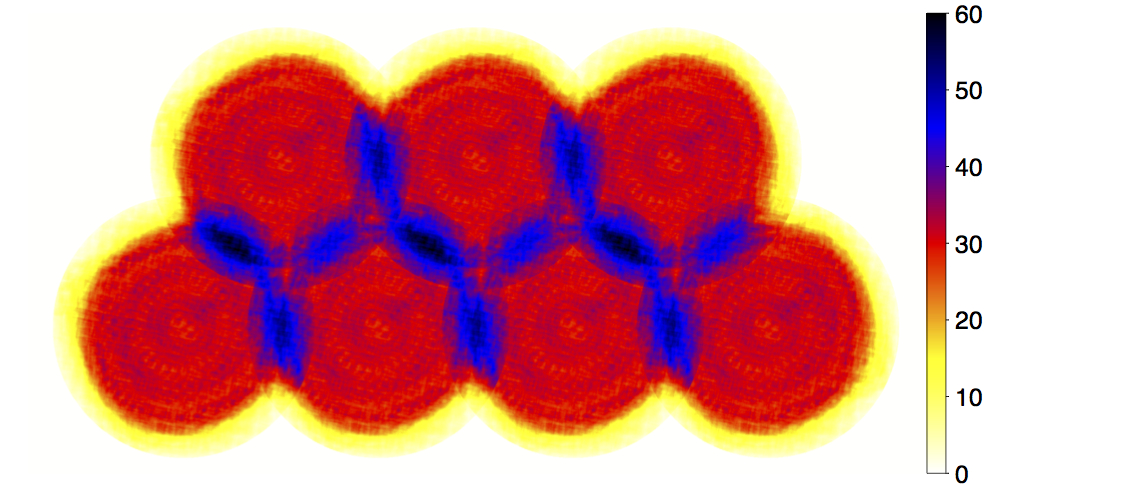}
\caption{Part of the survey grid pattern consisting of overlapping 1-degree circular {\em pong} tiles. The colour scale indicates the spatial distribution of relative exposure time (in arbitrary units) assuming identical observing conditions and the same number of repeats for each tile.
}
\label{grid}
\end{figure}

The survey target areas are sampled using a regular grid pattern of individual {\em pong3600} tiles with minimised overlaps.  Figure \ref{grid} shows an exposure-time map of part of such a grid.  The overlaps, in which the exposure time is roughly doubled, constitute about 9\% of the grid area.  Individual {\em pong3600}s have a pixel-to-pixel rms noise of 50 to 80 mJy per beam when reduced with 4-arcsecond pixels and each tile of the final survey will contain approximately seven repeats of these.  As far as possible, even coverage of the whole target area was maintained as the survey progressed, in order to produce relatively uniform sensitivity at any stage.

The 850-$\umu$m survey data presented in this paper cover the $\ell=30^{\rm o}$ field of the JPS and were observed between June 2012 and October 2013.  The 11 tiles making up the field were observed on average three times each.  A strategy of minimum elevation limits for given atmospheric opacity bands within the allocated range was adopted, in order to minimise variations in the resulting noise in each repeated tile.

Flux calibrations, pointing and focus checks were made routinely during each observing session with observations of isolated observatory standard point sources \citep{dempsey13}.  The flux calibrations were typically found to be repeatable to within 5\%.  The pointing accuracy is typically less than 5 arcseconds rms.

SCUBA-2 observes the 850-$\umu$m and 450-$\umu$m bands simultaneously but the assigned weather band for JPS observations ensures high sky opacity ($\tau \simeq 1.7$ to 2.7 in weather band 3) at 450\,$\umu$m.  The brightest sources are detected at low sensitivity at 450\,$\umu$m but the fluxes are not photometrically reliable and therefore are not currently included in the survey science data.

\section{Data Reduction}

The data presented in this paper were reduced with the Dynamic Iterative Map-Maker \citep{chapin13}, available as part of the {\em Starlink} {\sc smurf} package \citep{Jenness2011}.  

The raw data are first flat-fielded, converting raw DAC units into pW and then sampled to the requested pixel size, in this case 4 arcseconds, cleaned of spikes and steps in the data, and finally bad data are flagged or weighted down.

The iterative part of the reduction process then begins with the signal common to all bolometers being estimated and removed. An atmospheric correction is applied and the data are filtered to remove any remaining features on scales larger than the array footprint of 480 arcseconds, since these are likely to be artefacts. The remaining signal is then binned into a map of the sky using 4-arcsecond pixels. At this point the map made by the previous iteration (if any) is added back in, and the total map is sampled at the position of every bolometer value to create an estimate of the astronomical component of the signal, with any residual signal attributed to noise. These noise values are used to weight the bolometers when making maps on subsequent iterations, with each bolometer having multiple variance estimates calculated over a 15-second period. This astronomical signal is then removed from the original cleaned data, and the process is repeated iteratively starting from the modified data, until the change between maps produced by consecutive iterations is on average less than 1\% of the noise level.

%The above process divides each original bolometer value up into four additive components: common mode signal (COM), low frequency noise (FLT), astronomical signal (AST) and residual noise (RES). Using four components introduces a redundancy that allows signal to move between components in a poorly constrained manner. This lack of constraint allows strong low frequency structure to appear in the map (i.e.\ the AST component) provided that equal and opposite structures are added into the other components. These low frequency structures become stronger as more iterations are performed.  
In order to prevent the growth of low-frequency artefacts, an extra constraint is added by forcing all background pixels in the map to be zero. Background pixels are taken to be those with a signal-to-noise ratio less than 3.0, plus any other contiguously attached pixels down to a signal-to-noise ratio of 2.0. No such constraint is applied on the very last iteration, thus allowing non-zero values to appear in the background regions of the final map. This means that the background values within the final map are the result of a single iteration, while the astronomical signal has been arrived at by many iterations.

The present data have been reduced as individual {\em pong} tiles onto a 4-arcsecond pixel grid and then mosaiced to construct the whole $\ell=30^{\rm o}$ field map.  Prior to creating the mosaic, noisy tile edges were removed by cropping each observation to a 40 arcminute radius.  
The combined map was then lightly smoothed using a Gaussian kernel with $\sigma=1.8$ pixels (7.2 arcsec), approximating the half-width at half maximum of the telescope beam at 850\,$\umu$m.  
The standard observatory-determined calibration factor of 537\,Jy\,beam$^{-1}$\,pW$^{-1}$ was applied to the data.

\section{Results}

\begin{figure*}
\center
\includegraphics[width=0.95\textwidth,trim=0 0 0 0]{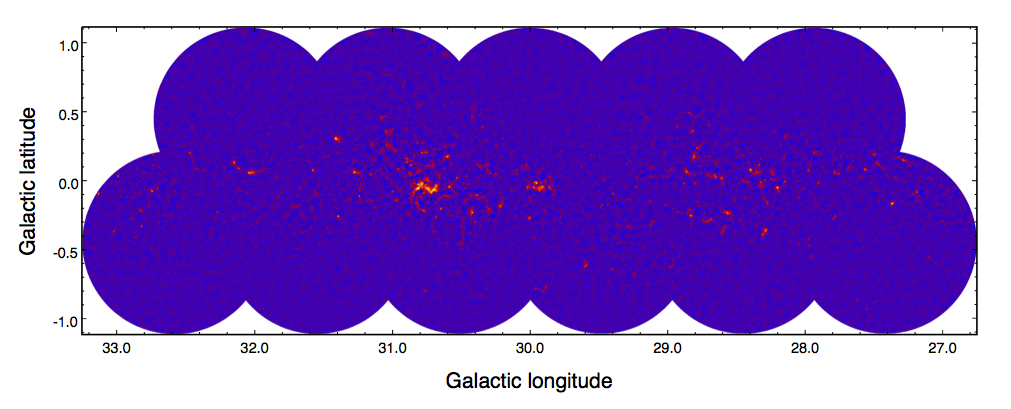}
\caption{JPS 850-$\umu$m surface-brightness distribution in the $\ell=30^\circ$ field as of October 2013 with about 40 per cent of the data acquired.  Individual circular pong tiles have been cropped at a radius of around 0.65$^{\circ}$.  The bright sources near longitudes of 30.0$^{\circ}$ and 30.8$^{\circ}$ are the well-known star-forming regions G29.96 and W43, respectively.  The colour scale range is logarithmic from zero to 23Jy\,beam$^{-1}$
}
\label{l30field}
\end{figure*}

\begin{figure*}
\center
\includegraphics[width=0.95\textwidth,trim=0 0 0 0]{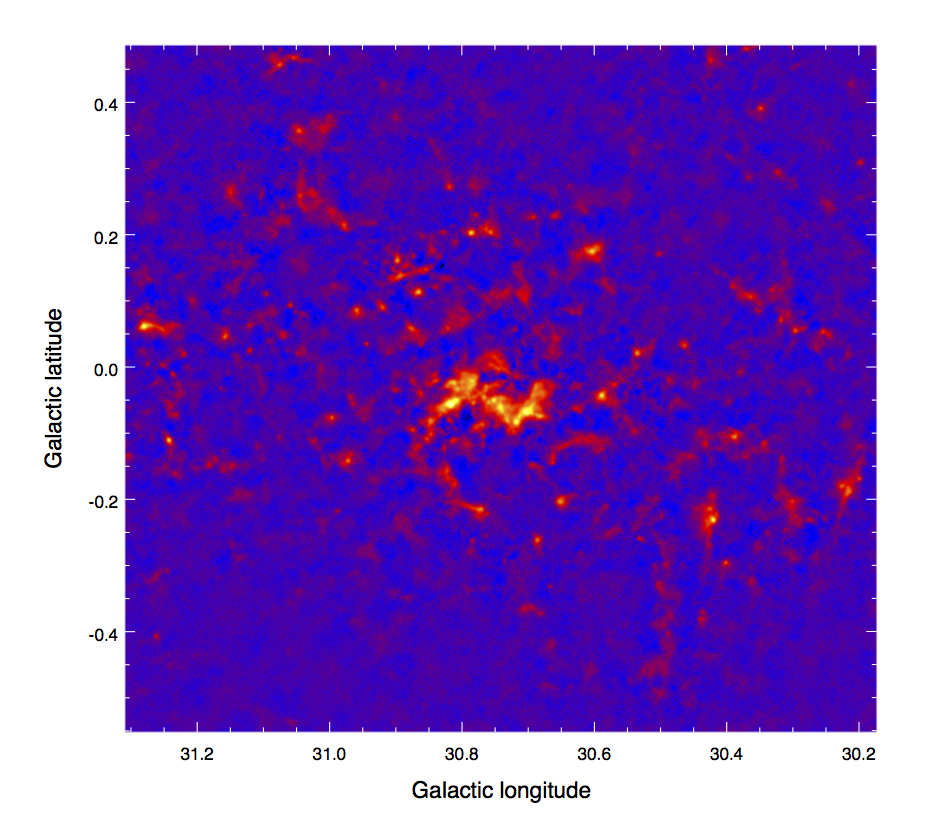}
\caption{850-$\umu$m surface-brightness distribution in the $\sim$1-degree square area around the W\,43 star-forming region.  The data have been lightly smoothed with a $\sigma = 7.2''$ Gaussian kernel, equivalent to the 850-$\umu$m JCMT beam, slightly degrading the native spatial resolution. 
}
\label{w43}
\end{figure*}

Figure \ref{l30field} shows all of the reduced $\ell=30^\circ$ field data.  Well-known star-forming regions W\,43 and G\,29.96 can be identified at positions $\ell=30.815$, $b=-0.057$ and $\ell=29.954$, $b=-0.018$, respectively.
Figure \ref{w43} shows a close-up of the same data in the 1-degree region around W\,43.

\begin{figure}
\includegraphics[width=0.47\textwidth, trim= 0 0 0 0]{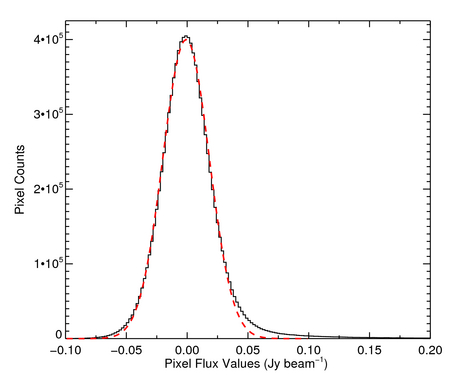}
\caption{The distribution of all pixel values in the $l=30^{\rm o}$ field (black histogram) and the result of a Gaussian fit to the distribution ($-$0.1 to 0.1\,Jy\,beam$^{-1}$; dashed red curve). The sensitivity of the current data is estimated from the latter fit to be $\sigma_{\rm{rms}} =19$\,mJy\,beam$^{-1}$.}
\label{fig:image_rms}
\end{figure}

The distribution of all pixel values in the $l=30^{\rm o}$ field is shown in Figure \ref{fig:image_rms}, along with a Gaussian fit to the distribution ($-$0.1 to 0.1\,Jy\,beam$^{-1}$). The rms noise in the data is estimated from the latter to be $\sigma_{\rm{rms}} =19$\,mJy\,beam$^{-1}$. It can be seen from the negative pixel values that there is a slight excess noise in the wings of the distribution above that of the assumed Gaussian.  The larger excess on the positive side is due to signal from real sources.  Since the data set used here consists of around 40 per cent of the eventual JPS data, the projected rms sensitivity in the complete survey, in regions where confusion is not dominant, is therefore likely to be around 12\,mJy\,beam$^{-1}$.  This can be compared to the ATLASGAL rms sensitivity of 50 -- 80\,mJy.

\subsection{$^{12}$CO $J$=3-2 line contamination}

\begin{figure*}
\includegraphics[width=0.95\textwidth, trim= 0 0 0 0]{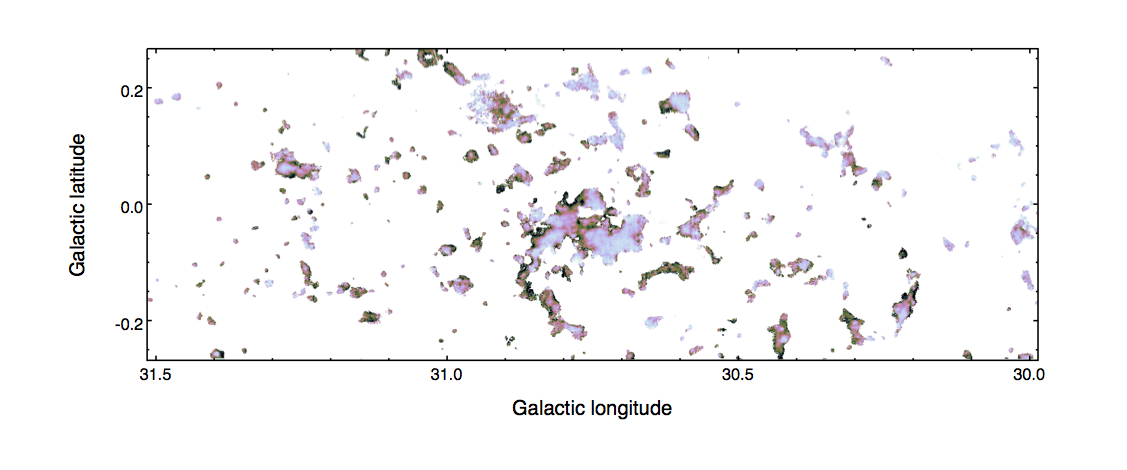}
\caption{Spatial distribution of the $^{12}$CO J=3--2 emission line contamination in the region around W43.  The colour scale shows the value of $(S(850) - S(^{12}\mbox{CO}))/S(850)$ and the range is from 0.7 (darkest) to 1.0 (lightest).  
}
\label{COcorrmap}
\end{figure*}

With a central frequency of 355\,GHz and half-power width around 35\,GHz, the continuum bandpass at 850\,$\umu$m will suffer contamination from spectral-line emission, predominantly by $^{12}$CO ($J$=3-2) at 345.796\,GHz \citep*{johnstone03}. To quantify the degree of contamination, we reprocess the SCUBA-2 data with the CO contribution removed during the reduction.

CO data for this region were taken from the CO High Resolution Survey (COHRS: \citealp*{dempsey13a}). COHRS has mapped the Galactic plane in $^{12}$CO ($J$=3-2) in the Galactic longitude range $\ell = 10^{\rm o}$ to $\ell=55^{\rm o}$ using HARP on the JCMT and its spatial resolution is therefore essentially identical to that of the SCUBA-2 data. COHRS only extends to Galactic latitudes $b=\pm 0.25^\circ$ in the $\ell = 30^{\circ}$ region, so only this central strip can be considered in this analysis.

The technique for removing the CO contamination involves running the mapmaker as described in Section 3 but with the CO map, converted to SCUBA-2 output power units, supplied as a negative fake source. This method has previously been used by \cite{drabek12}, \cite{hatchell13} and \cite{sadavoy13} 
%and \cite{parsons14} 
to investigate CO contamination. To convert to physical units (pW), we begin by multiplying the CO intensities (in K\,km\,s$^{-1}$) by the forward-scattering and spillover efficiency (0.71) and by a molecular-line conversion factor. This factor is calculated from the atmospheric precipitable water vapour ($P$) and scales with opacity at 225\,GHz using the following equation, described in full in %\citep{parsons14}:
Parsons et al.\ (in preparation):

\begin{align*}
\noindent
%\[
%\text{C}_{850}\text{ = } & 0.57417 + 0.115089 \text{\,PWV} - 0.048470 \text{\,PWV}^2 \nonumber \\
%& + 0.010893 \text{\,PWV}^3 - 0.000856 \text{\,PWV}^4  \nonumber \\
%& \text{mJy\,beam}^{-1} / \text{(K\,km\,s}^{-1}\text{)}
C_{850} & = 0.57417 + 0.115089 \,P - 0.048470 \,P^2 
+ 0.010893 \,P^3 \\ & - 0.000856 \,P^4 \,
\text{mJy\,beam}^{-1} / \text{(K\,km\,s}^{-1}\text{)}
\label{eq:C850}
%\]
\end{align*}

The final step converts the CO data from mJy\,beam$^{-1}$ into negative pW by dividing by the standard 850-$\umu$m calibration factor ($-$537000 mJy\,beam$^{-1}$\,pW$^{-1}$). Once the raw data have been reduced with the CO contribution removed, the resulting map is compared to the uncorrected SCUBA-2 data, applying a mask based on a 3-$\sigma$ cut in the signal-to-noise map of the original reduction to both maps.

For the bulk of the data, the levels of CO contamination range from zero to $\sim$30\% (see Figure \ref{COcorrmap}) and the majority of values are only a few per cent. This is consistent with values reported by \cite{schuller09} who quote results from \cite{wyrowski06} and with those found by \cite{drabek12}.
The distribution is not random but varies coherently between and within sources, with clearly resolved gradients that are sometimes radial and sometimes longitudinal.  In general, the highest levels of contamination appear towards the edges of continuum sources, where the $^{12}$CO optical depth is likely to be least.  Detailed analysis of the spatial variation in CO contamination is beyond the scope of this paper and will be addressed in a future study; however, for the brightest sources with a SCUBA-2 flux between 1000 and 1500\,mJy\,beam$^{-1}$, the contamination levels have a mode of 2\% and do not exceed 10\%. It is therefore clear that the level of contamination is relatively low and unlikely to affect significantly the extracted dust column densities.  

Since the JPS has a wider latitude range ($|\,b\,| \le 0.85^{\rm o}$) than COHRS, a statistical correction would be desirable where no COHRS data are available.  However, we found no correlation between the level of contamination and the surface brightness in the continuum map and so such an extrapolation is unlikely to be possible.

Another potential source of contamination of the thermal dust emission at sub-millimetre wavelengths is that from the free-free continuum.  Obviously, this would be most severe for luminous compact H{\sc ii} regions, such as those in W43.  \cite{schuller09}, using results from \cite{motte03}, estimate this contribution to be less than $\sim$20 per cent in most cases.

\subsection{Detection of compact sources}

To identify compact sources within the data, we have used the \FW\ (FW; \citealp{berry15}) source-extraction algorithm, which is part of the {\em Starlink} software package {\sc cupid} (\citealp{berry07}). \cite{berry15} shows that FW is robust against a wide choice of input parameters, which is not the case for Clumpfind or Gaussclumps \citep{watson10}.  An overview of these and other algorithms can be found in \cite{menshchikov}.  Like most source extraction algorithms, FW performs 
best when the background noise is uniformly distributed. We have therefore used the weight map produced by the data-reduction processing to estimate the rms noise distribution in the image and so produced a signal-to-noise ratio (SNR) map of the whole region. The noise in this SNR map is effectively constant, even towards the edges where the rms in the surface-brightness map increases significantly. 

The SNR map was used as the initial input into FW and all sources above a specified SNR threshold were identified with the task {\sc{Findclumps}}.  The FW algorithm assigns pixels to clumps by ascending towards emission peaks from each pixel above the threshold via the steepest gradient.  It then creates a mask of the input map in which pixel values either identify the source to which the pixel has been assigned or are set to a negative value if unassigned. This mask is then used as input to the task {\sc{Extractclumps}} that extracts peak and integrated flux values from the original emission map for each source identified. 

We set a detection threshold of 3$\sigma$ and required that the source consisted of $>$7 pixels; this is the number of pixels we would expect above the detection threshold for an unresolved 5-$\sigma$ source.  The sensitivity of JPS data to the extended, filamentary structures found to be ubiquitous in {\em Herschel} Hi-GAL images (\citealp{molinari10b}) has not yet been tested and so, for the purposes of this paper, we have rejected sources with aspect ratios larger than 5.

\begin{figure*}
\center
\includegraphics[width=0.95\textwidth, trim = 0 0 0 0]{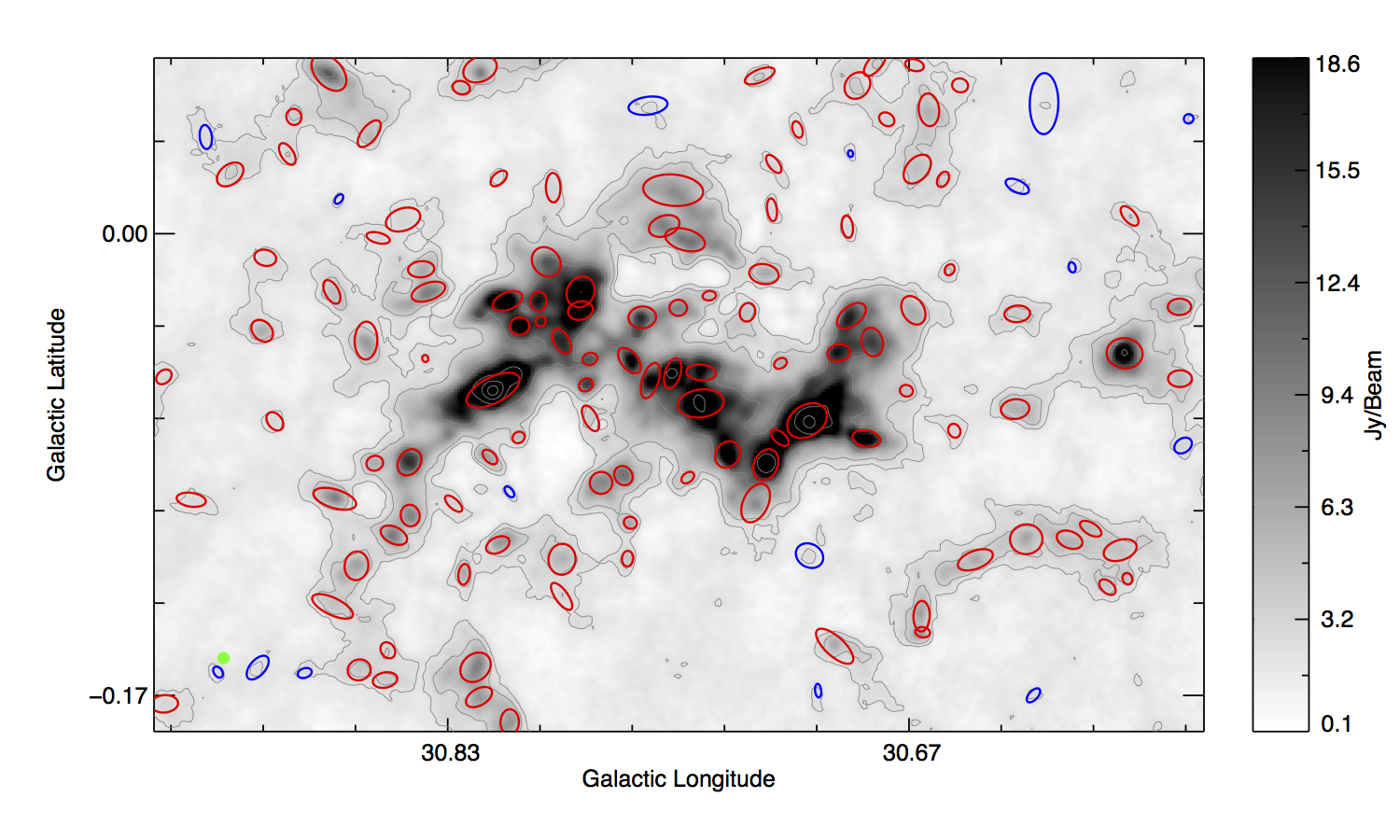}
\caption{Example of the source-extraction results in a small portion of the $\ell = 30\degr$ region. The greyscale image shows the 850-$\umu$m emission and the red and blue ellipses show the approximate sizes and orientations of the sources above and below 5$\sigma$. The grey contours start at 3$\sigma$ and increase in steps of 0.17, 0.8, 2.75, 6.81, 13.85\, Jy\,beam$^{-1}$. The green filled circle in the lower-left corner indicates the FWHM size of the JCMT beam at 850\,$\umu$m.}
\label{fig:source_extraction_example}
\end{figure*}

Figure\,\ref{fig:source_extraction_example} displays the distribution of 850-$\umu$m surface brightness in a small subregion of the $l=30^{\rm o}$ field, around the prominent W43 star-forming region, and the sources extracted from the data by FW.  Most of the sources that are identifiable by eye have been recovered, although the detection of sources below 5$\sigma$ is relatively unreliable, further justifying a threshold at this level for inclusion in our preliminary compact source catalogue.  The latter consists of 1029 objects and is listed in Table \ref{table:cattable}.  The set of FW parameter settings used is given in Table \ref{table:fwtable}

\subsection{Flux calibration and photometric corrections}

\FW\ produces peak flux density values taken simply from the highest pixel value associated with each extracted source.  Integrated flux densities are calculated by summing the values of all associated pixels in the source mask and dividing the result by the so-called beam integral, which is the number of pixels per beam.  The latter is determined from the ratio of integrated to peak surface brightness of a point source.  In this case, the observations of Neptune, reduced 
in the same manner as the science data were used, producing a beam integral value of 
17.5, which is significantly larger than that expected for a Gaussian beam ($\sim$14), although the FWHM of the Neptune image is consistent with the latter.  This is due to significant power in the wings of the JCMT beam, as discussed by \citet{dempsey13}.

\begin{figure}
\includegraphics[width=0.47\textwidth, trim= 0 0 0 0]{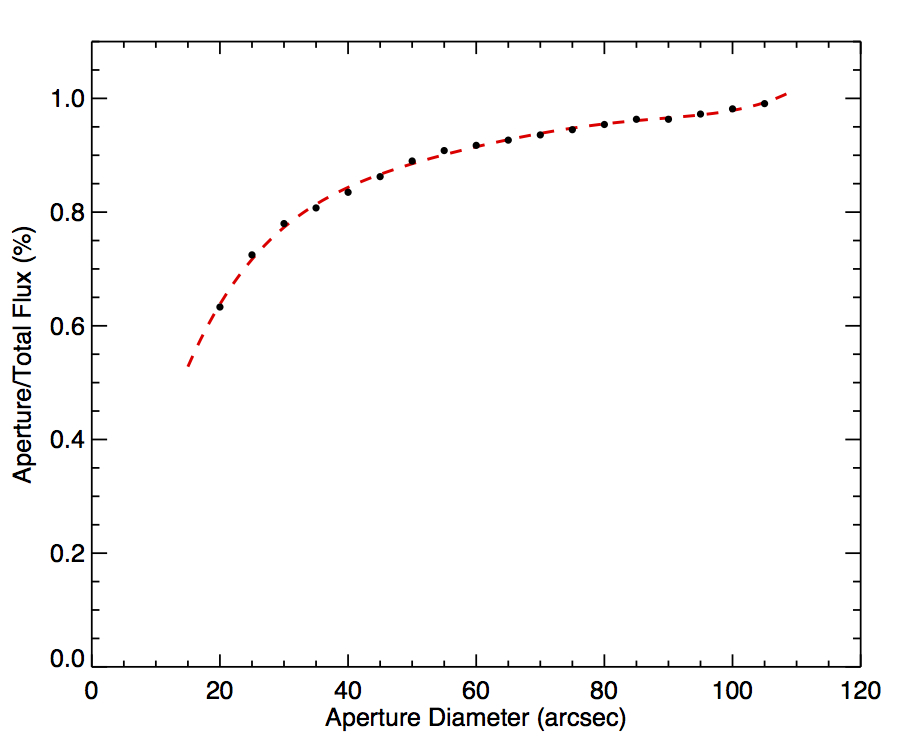}
\caption{Aperture corrections determined by \citet{dempsey13} (filled circles) and re-normalised to a value of 100\% at the largest aperture, then fitted with a fifth-order polynomial fit (dashed line), from which corrections were applied to the integrated fluxes produced by \FW.}
\label{apcorr}
\end{figure}

In addition to this, the FW method, in common with similar algorithms, effectively uses variable aperture photometry to produce integrated flux densities, with the aperture size set by the number of associated pixels above the inclusion threshold in the source masks.  This number is SNR-dependent and so requires the application of an aperture correction.  This has been done by converting the area in pixels ($A$) of each source to an effective source circular radius $r = \theta \sqrt{A/\pi}$, where $\theta$ is the angular pixel scale, and then using the aperture correction values tabulated by \citet{dempsey13}.  The latter were checked using our own observation of Neptune, and found to be consistent to within 1 per cent.  Values intermediate to those tabulated were interpolated using a polynomial fit (see Figure \ref{apcorr}). Note that these corrections are only approximate, since they take no account of departures from circular symmetry.   However, they only significantly affect small sources, those with sizes only a little larger than the beam, and these are likely to be circular.  Most irregular sources are extended with respect to the beam and for sources with diameters $> 60$ arcseconds the correction is of order 10 per cent or less.  It is also the case that very compact and unresolved sources are relatively rare.
The error beam contribution is estimated to have an amplitude of 2 per cent \citep{dempsey13}, which is much less than the other uncertainties.

\begin{table*}
\begin{center}
\caption{ The JPS $\ell=30^{\rm o}$ preliminary compact source catalogue.  The columns are as follows: (1) name derived from Galactic coordinates of the maximum intensity in the source; (2)-(3) Galactic coordinates of maximum intensity in the catalogue source; (4)-(5) Galactic coordinates of emission centroid; (6)-(8) semi-major and semi-minor size and source position angle measured anti-clockwise from Galactic north; (9) effective radius of source, i.e., $R_{\rm eff}=\sqrt{\left(A/\pi\right)}$, where $A$ is the area of the source above the threshold; (10)-(13) peak and integrated flux densities and their associated uncertainties; (14) signal to noise ratio (SNR).}
\label{table:cattable}
\begin{tabular}{lllllrrrrrrrrr}
  \hline \hline
  \multicolumn{1}{c}{Name} 
  &  \multicolumn{1}{c}{$\ell_{\mathrm{max}}$} &  \multicolumn{1}{c}{$b_{\mathrm{max}}$}
  &  \multicolumn{1}{c}{$\ell$} &  \multicolumn{1}{c}{$b$} &
  \multicolumn{1}{c}{$\sigma_{\rm{maj}}$} &  \multicolumn{1}{c}{$\sigma_{\rm{min}}$} &  \multicolumn{1}{c}{PA} &
  \multicolumn{1}{c}{$R_{\rm{eff}}$} &  \multicolumn{1}{c}{$S_{\rm{peak}}$} & \multicolumn{1}{c}{$\Delta S_{\rm{peak}}$} & \multicolumn{1}{c}{$S_{\rm{int}}$}& \multicolumn{1}{c}{$\Delta S_{\rm{int}}$} &  \multicolumn{1}{c}{SNR} \\  
  \multicolumn{1}{c}{} &  \multicolumn{1}{c}{($^{\circ}$)} &
  \multicolumn{1}{c}{($^{\circ}$)} &  \multicolumn{1}{c}{($^{\circ}$)} &
  \multicolumn{1}{c}{($^{\circ}$)} &  \multicolumn{1}{c}{($''$)}
  &  \multicolumn{1}{c}{($''$)} &  \multicolumn{1}{c}{($^{\circ}$)}&  \multicolumn{1}{c}{($''$)}
  &  \multicolumn{2}{c}{(Jy\,beam$^{-1}$)} &  \multicolumn{2}{c}{(Jy)}&\\
  \multicolumn{1}{c}{(1)} &  \multicolumn{1}{c}{(2)} &  \multicolumn{1}{c}{(3)} &  \multicolumn{1}{c}{(4)} &
  \multicolumn{1}{c}{(5)} &  \multicolumn{1}{c}{(6)} &  \multicolumn{1}{c}{(7)} &  \multicolumn{1}{c}{(8)} &
  \multicolumn{1}{c}{(9)} &  \multicolumn{1}{c}{(10)} &  \multicolumn{1}{c}{(11)} & \multicolumn{1}{c}{(12)} &  \multicolumn{1}{c}{(13)} &  \multicolumn{1}{c}{(14)}  \\
  \hline
G026.830$-$00.208	&	26.830	&	$-$0.208	&	26.830	&	$-$0.206	&	7	&	6	&	259	&	16	&	0.38	&	0.08	&	0.64	&	0.13	&	5.6	\\
G026.865$-$00.276	&	26.865	&	$-$0.276	&	26.867	&	$-$0.276	&	12	&	11	&	112	&	26	&	0.36	&	0.06	&	1.00	&	0.17	&	6.8	\\
G026.957$-$00.077	&	26.957	&	$-$0.077	&	26.956	&	$-$0.076	&	10	&	8	&	224	&	26	&	0.77	&	0.08	&	1.64	&	0.17	&	15.2	\\
G026.960$-$00.309	&	26.960	&	$-$0.309	&	26.962	&	$-$0.308	&	14	&	7	&	177	&	23	&	0.20	&	0.04	&	0.67	&	0.14	&	5.0	\\
G027.000$-$00.298	&	27.000	&	$-$0.298	&	27.001	&	$-$0.297	&	24	&	11	&	167	&	38	&	0.69	&	0.07	&	1.89	&	0.20	&	15.1	\\
G027.011$-$00.040	&	27.011	&	$-$0.040	&	27.012	&	$-$0.039	&	11	&	11	&	151	&	30	&	0.47	&	0.06	&	1.15	&	0.14	&	10.4	\\
G027.019$-$00.166	&	27.019	&	$-$0.166	&	27.021	&	$-$0.167	&	13	&	6	&	192	&	21	&	0.20	&	0.04	&	0.51	&	0.11	&	5.1	\\
G027.037$-$00.171	&	27.037	&	$-$0.171	&	27.035	&	$-$0.165	&	22	&	12	&	269	&	31	&	0.22	&	0.04	&	0.98	&	0.19	&	5.6	\\
G027.066+00.013	&	27.066	&	$+$0.013	&	27.078	&	$+$0.015	&	32	&	16	&	163	&	44	&	0.25	&	0.05	&	1.76	&	0.35	&	5.4	\\
G027.086$-$00.564	&	27.086	&	$-$0.564	&	27.086	&	$-$0.563	&	9	&	8	&	149	&	22	&	0.39	&	0.05	&	0.92	&	0.13	&	8.7	\\
G027.100$-$00.559	&	27.100	&	$-$0.559	&	27.098	&	$-$0.557	&	17	&	8	&	186	&	30	&	0.57	&	0.06	&	1.48	&	0.17	&	12.9	\\
G027.173$-$00.008	&	27.173	&	$-$0.008	&	27.173	&	$-$0.005	&	14	&	10	&	210	&	27	&	0.27	&	0.05	&	0.83	&	0.14	&	6.5	\\
G027.178$-$00.105	&	27.178	&	$-$0.105	&	27.176	&	$-$0.105	&	12	&	9	&	166	&	27	&	0.43	&	0.06	&	1.29	&	0.17	&	9.4	\\
G027.186$-$00.082	&	27.186	&	$-$0.082	&	27.186	&	$-$0.079	&	29	&	16	&	268	&	55	&	3.07	&	0.25	&	6.19	&	0.50	&	67.8	\\
G027.187$-$00.148	&	27.187	&	$-$0.148	&	27.187	&	$-$0.148	&	20	&	11	&	213	&	33	&	0.23	&	0.05	&	1.15	&	0.23	&	5.6	\\
G027.207$-$00.100	&	27.207	&	$-$0.100	&	27.208	&	$-$0.098	&	16	&	15	&	214	&	35	&	0.44	&	0.06	&	2.09	&	0.28	&	9.6	\\
G027.221+00.136	&	27.221	&	$+$0.136	&	27.221	&	$+$0.136	&	16	&	8	&	188	&	30	&	0.84	&	0.09	&	1.70	&	0.18	&	14.3	\\
G027.248+00.108	&	27.248	&	$+$0.108	&	27.245	&	$+$0.108	&	16	&	14	&	189	&	34	&	0.53	&	0.07	&	1.64	&	0.21	&	10.4	\\
G027.256+00.135	&	27.256	&	$+$0.135	&	27.255	&	$+$0.136	&	10	&	7	&	130	&	22	&	0.68	&	0.08	&	1.34	&	0.15	&	11.9	\\
G027.279+00.145	&	27.279	&	$+$0.145	&	27.272	&	$+$0.147	&	30	&	12	&	193	&	46	&	1.66	&	0.15	&	6.06	&	0.53	&	28.5	\\
  \hline
\end{tabular}
\end{center}
Note: Only a small portion of the data is provided here, the full table is only  available in electronic form at the CDS via anonymous ftp to cdsarc.u-strasbg.fr (130.79.125.5) or via http://cdsweb.u-strasbg.fr/cgi-bin/qcat?J/%{\color{red}????}/.
\end{table*} 

\begin{table}
\begin{center}
\caption{\FW\ parameter settings used in the source extraction
}
\label{table:fwtable}
\begin{tabular}{ll}
  \hline \hline
\sc fellwalker.allowedge & 0 \\
\sc fellwalker.cleaniter & 5 \\
\sc fellwalker.flatslope & 1 \\
\sc fellwalker.fwhmbeam & 1 \\
\sc fellwalker.maxbad & 0.05 \\
\sc fellwalker.maxjump & 3 \\
\sc fellwalker.mindip & 1.5 \\
\sc fellwalker.minheight & 3 \\
\sc fellwalker.minpix & 7 \\
\sc fellwalker.noise & 1 \\
\sc fellwalker.rms & 1 \\
\sc fellwalker.shape & ellipse \\
  \hline
\end{tabular}
\end{center}
\end{table} 

\begin{figure}
\center
\includegraphics[width=0.47\textwidth, trim= 0 0 0 0]{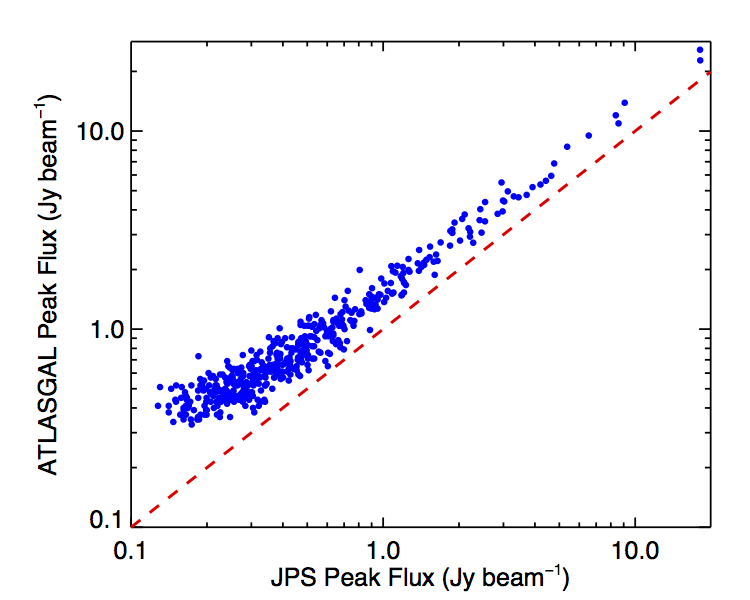}
\includegraphics[width=0.47\textwidth, trim= 0 0 0 0]{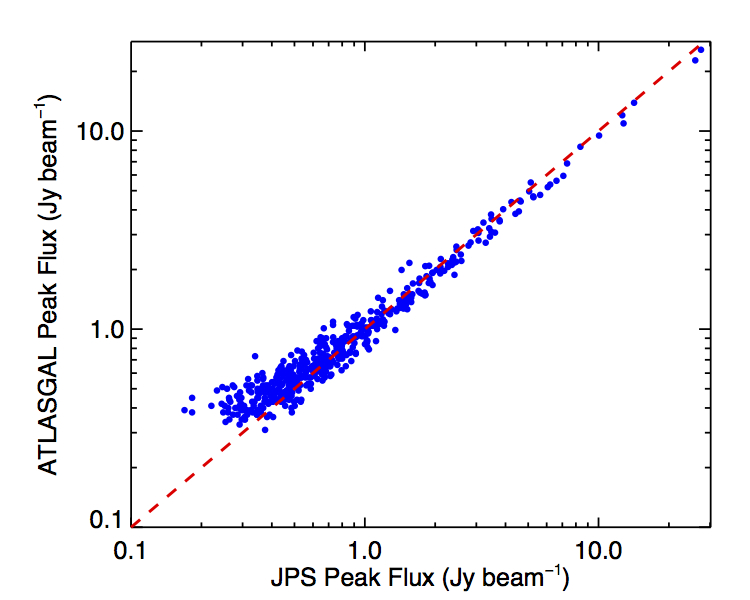}
\caption{Comparison of the peak flux densities of extracted JPS sources to the corresponding ATLASGAL compact sources - top: unsmoothed data with 484 matches; bottom: with JPS data convolved to the resolution of ATLASGAL, producing 500 matches.
}
\label{JPS_ATLAS}
\end{figure}

As a check on the photometry of the sources extracted by \FW, we have compared the peak flux densities of the JPS compact sources to those of spatially corresponding detections in the ATLASGAL survey \citep{contreras13}.  We searched the ATLASGAL compact-source catalogue \citep{urquhart14b} for the closest spatial coincidence within 10 arcseconds of each JPS detection and found 484 matching sources.  Figure \ref{JPS_ATLAS} (upper panel) shows the comparison of peak flux densities for this initial matched list.  It can be seen that the JPS peak flux densities are systematically lower than those of ATLASGAL, with a scatter that increases towards lower values.  Much of this systematic difference can be attributed to the fact that most sources in the ATLASGAL catalogue were found to be at least somewhat extended.  Given the larger APEX beam (19 arcsec, compared to around 14.5 arcsec for JCMT) at these wavelengths, we expect peak flux densities to be larger in the former data.  To correct for this, we repeated the JPS source extraction after convolving the data to the ATLASGAL spatial resolution and scaling the signal level so as to restore the (equality of) peak and integrated flux densities of point sources.  The lower panel of Figure \ref{JPS_ATLAS} shows the result, this time with 500 positional matches and a significant improvement in the flux matching.  The mean ratio of peak flux densities in the two catalogues for these latter matched sources is $S_{850}$(JPS)/$S_{870}$(ATLAS) = $0.93 \pm 0.01$ and a standard deviation of 0.12.  Despite being observed with different telescopes and detectors and using unrelated techniques, reduced using different procedures and the source extraction being done using independent methods (ATLASGAL uses {\sc Sextractor}), the fluxes in the two catalogues are in very good agreement.  This is, therefore, an important consistency check on the reliability of the results of both catalogues.

The scatter in the correlation widens at lower flux densities as might be expected where the signal-to-noise ratio in the ATLASGAL data becomes low.  However, there is also a deviation from the correlation at the low end, in the form of a tail with high values of $S_{850}$(JPS)/$S_{870}$(ATLAS). This tail may be the result of a bias related to the flux-boosting bias found in the JPS data (see Section 4.5).  Without it, the correspondence between the two flux scales becomes almost perfect.  Taking only JPS fluxes above 0.5 Jy\,beam$^{-1}$, the mean peak-flux-density ratio in the matched sample is $1.001 \pm 0.001$ and the median value is 1.012.

\subsection{Completeness tests}

To test the efficiency of the source-extraction algorithm as a function of source flux we injected artificial compact sources onto a field of uniform noise, the latter distributed as a Gaussian with a standard deviation of $\sim$20\,mJy\,beam$^{-1}$, similar to that in the data. The source profiles were Gaussian and circularly symmetrical with a FWHM of 5 pixels (20 arcsec).  \FW\ was then used with the same settings as in the survey data to recover the artificial sources.  The rate of detections as a function of input flux density then allows us to benchmark the performance of the extraction algorithm in this idealised case.

\begin{figure}
\includegraphics[width=0.47\textwidth, trim= 0 0 0 0]{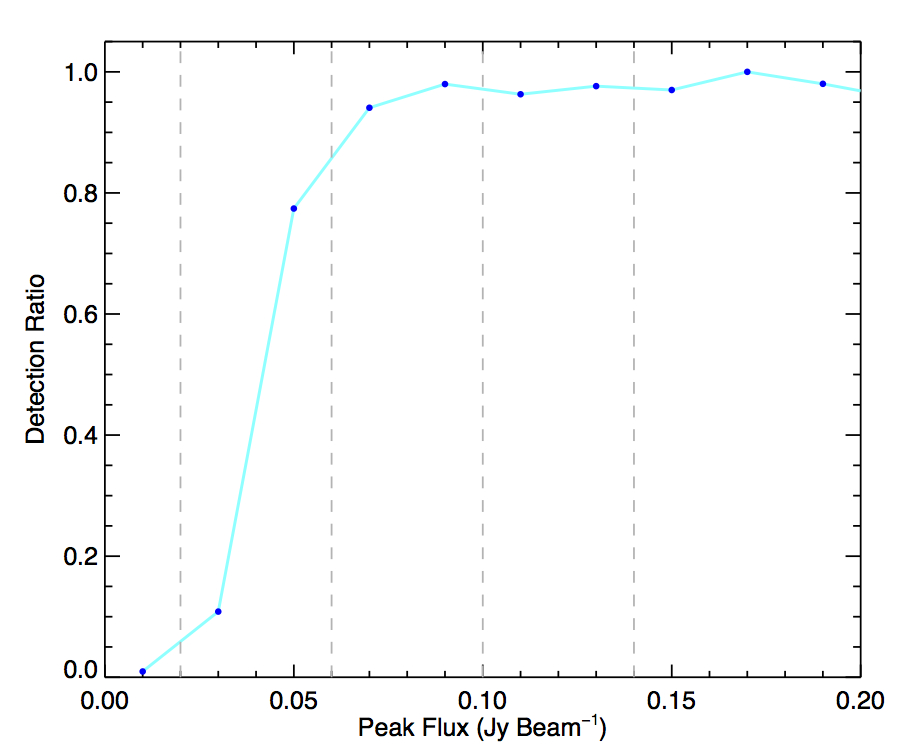}
\caption{The ratio of recovered to injected source counts as a function of input peak flux density. The dashed vertical lines indicate (from left to right) the 1$\sigma$, 3$\sigma$, 5$\sigma$, and 7$\sigma$ noise levels.}
\label{completeness}
\end{figure}

The {\sc cupid} task {\sc Makeclumps} was used to generate a catalogue of
artificial objects with a uniform distribution in $\ell$, $b$. The peak fluxes of the
injected objects were uniformly distributed between 0.02 and
0.5\,Jy\,beam$^{-1}$ and the source profiles were Gaussian with FWHM of 5 pixels
(20\arcsec). We used the size of the survey field as a template and injected a similar number of sources as was detected by \FW\ in the data.  Figure\,\ref{completeness} shows the number of sources recovered as a function of input peak flux and that the source extraction algorithm is approximately 95\% complete above 80\,mJy\,beam$^{-1}$ ($\sim$4$\sigma$).

\begin{figure}
\includegraphics[width=0.47\textwidth, trim= 0 0 0 0]{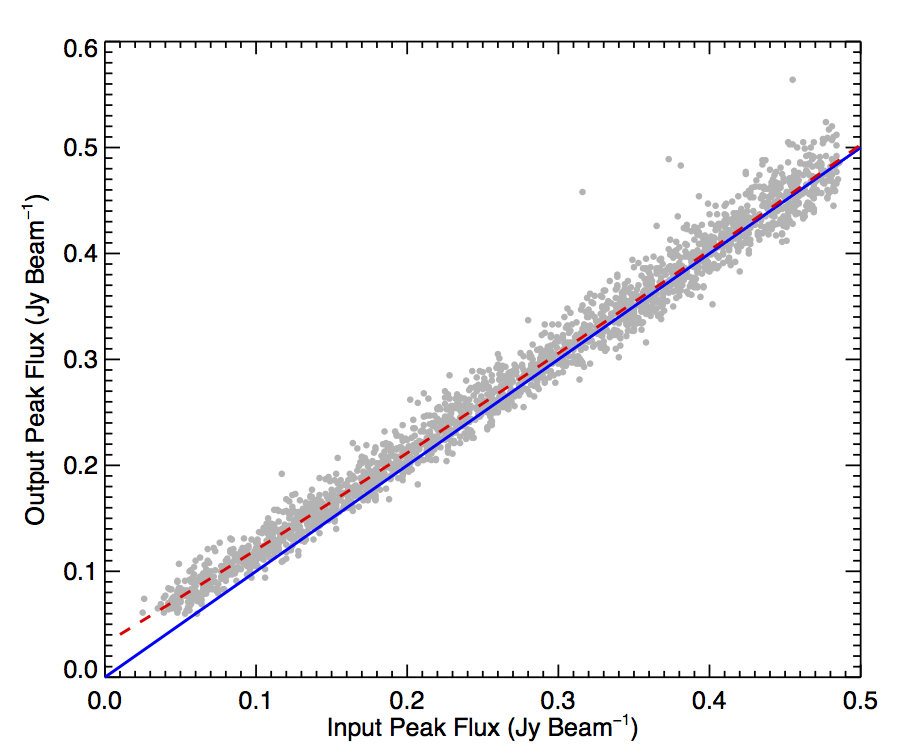}
\caption{A comparison of recovered peak flux densities to the input values for the artificial sources used in the completeness tests. The solid blue line represents equality and the dashed red curve shows the result of a polynomial fit to the data.}
\label{input_output_comparison}
\end{figure}

Figure\,\ref{input_output_comparison} compares the values of the original and recovered peak flux densities of the injected sources.  These are very well correlated, showing that, in general, source fluxes are recovered accurately. However, the recovered fluxes for the weaker sources are systematically higher than their injected values. The reason for this is that \FW\ sets the peak flux to be the value of the brightest pixel in the source and, since the injected sources are all slightly extended, there is often a pixel with a positive noise deviation near the peak that adds constructively to the measured peak flux, creating a small bias. This results in the average measured flux density being approximately 1$\sigma$ larger than expected for the weaker sources. This systematic bias to higher flux values for weaker sources is well fitted by a second order polynomial function (Figure\,\ref{input_output_comparison}) and the recovered real source flux densities can be corrected accordingly.

The polynomial coefficients are: $S_\nu({\rm{corr}}) = -0.0374 + 1.187\,S_\nu({\rm{orig}}) - 0.265\,S_\nu({\rm{orig}})^2$, where $S_\nu({\rm{corr}})$ and $S_\nu({\rm{orig}})$ are the corrected and original peak flux values (in Jy\,beam$^{-1}$), respectively. We apply this correction to all sources with peak flux densities below 0.3 Jy\,beam$^{-1}$.  Above this value the correction is not necessary as it is only a few percent of the flux value, significantly less than the $\le 5$ percent flux calibration uncertainty in 850-$\umu$m SCUBA-2 data.

%The polynomial coefficients are: $S_\nu({\rm{corr}}) = -0.0373905 + 1.18751 \times\ S_\nu({\rm{orig}}) - 0.265127 \times S_\nu({\rm{orig}})^2$, where $S_\nu({\rm{corr}})$ and $S_\nu({\rm{orig}})$ are the corrected and measured peak flux values respectively. We apply this correction to all sources with measured flux value below 0.3 Jy\,beam$^{-1}$; above this value the correction is not necessary.

\subsection{Flux distributions (peak and integrated)}

\begin{figure}
\includegraphics[width=0.47\textwidth, trim= 0 0 0 0]{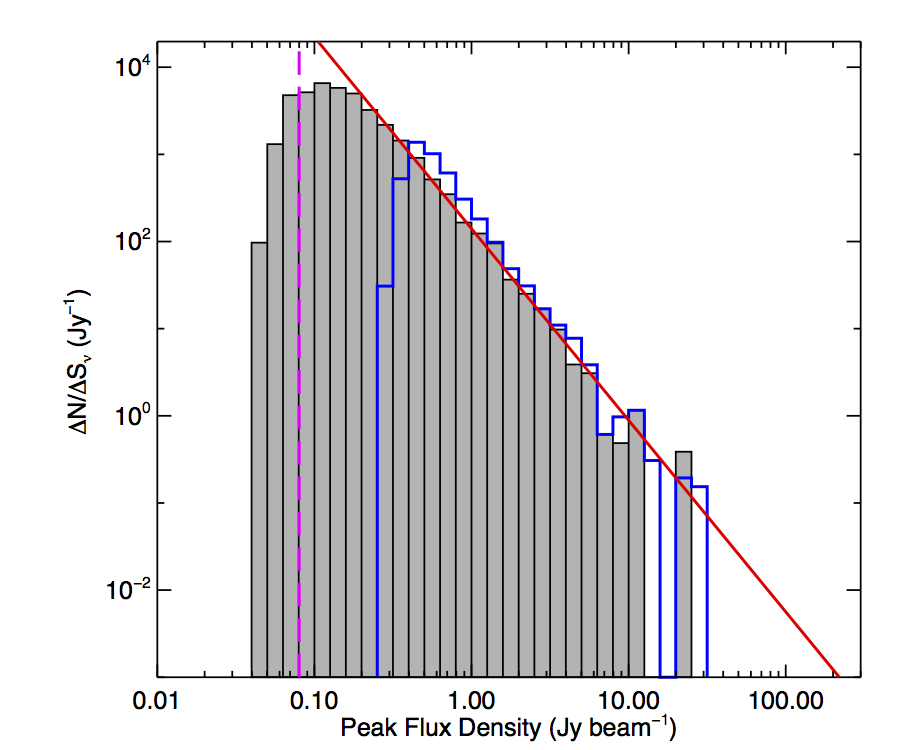}
\caption{Distribution of peak flux densities of all extracted compact sources in the $\ell=30$ field (grey shaded histogram) with the equivalent ATLASGAL distribution in the same projected area of sky (blue outline histogram).  The red line indicates the linear least-squares fit to the data above the completeness turnover.
}
\label{peakfd}
\end{figure}

\begin{figure}
\includegraphics[width=0.47\textwidth, trim= 0 0 0 0]{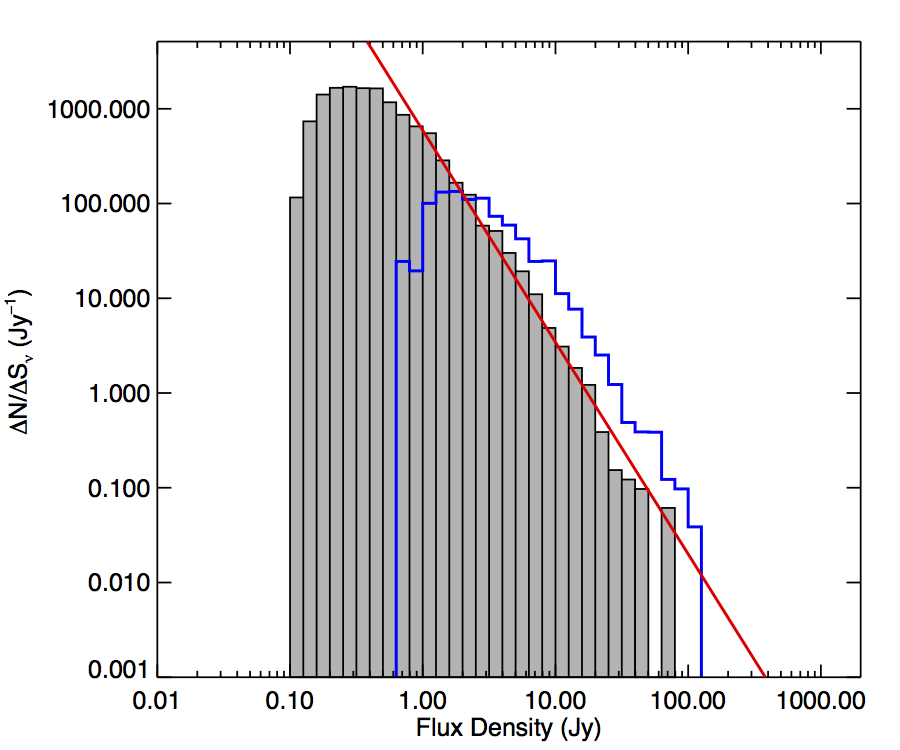}
\caption{Distribution of integrated flux densities of the extracted compact sources in the $\ell=30$ field  (grey shaded histogram) with the equivalent ATLASGAL distribution in the same projected area of sky (blue outline histogram)}
\label{integfd}
\end{figure}

Figure \ref{peakfd} shows the distribution of peak flux densities in the sample of compact sources extracted from the $\ell = 30$ field, as described above and listed in Table 1.  Similarly, Figure \ref{integfd} shows the distribution of integrated flux densities.  Both of these are compared, without corrections for differing beam size, to the equivalent distributions in the sources extracted from the ATLASGAL compact-source catalogue  (\citealp{contreras13, urquhart14b}; note that a later ATLASGAL catalogue is available: \citealp{csengeri14}) in the same area of sky.  Both figures include a power-law fit to the JPS data above the turnovers in each case.

The distributions in Figure \ref{peakfd} are very similar, with much the same slope that approximates well to a power law in both data sets.  The main difference is that the ATLASGAL distribution turns over at a peak flux density about five times higher than that of the JPS sample and has a slight excess of brighter sources.  In contrast, the integrated flux-density distributions in Figure \ref{integfd} are significantly different. The ATLASGAL sample distribution is shifted to values that are around a factor of two higher than for JPS and appears to be less well represented by a power law.  The latter difference is likely to be the result of source merging in the larger APEX telescope beam ($\sim$19$''$ FWHM).

The turnover in the peak flux-density distribution, which can normally be assumed to mark the effects of incompleteness in the sample, is significantly higher than the limit calculated from the tests on artificial data described above.  This is likely to be because of confusion due to source crowding and perhaps non-uniform noise, effects that are likely to be position-dependent within the survey.  If this is the case, the true completeness limit in the $\ell = 30^{\circ}$ field may be as high as 0.1--0.2\,Jy\,beam$^{-1}$ (which may be compared to the corresponding limit in the ATLASGAL survey at  0.3--0.5\,Jy\,beam$^{-1}$, \citealp{contreras13}).  Estimating again from the turnovers in the distributions, the JPS integrated source flux densities appear complete above $\sim$0.8\,Jy while ATLASGAL limit is around 2--3\,Jy.  More rigorous tests will be conducted to quantify completeness when the entire JPS data set is available.

\subsection{The Mass Distribution of Clumps in W\,43}

\begin{figure}
\includegraphics[width=0.47\textwidth, trim= 0 0 0 0]{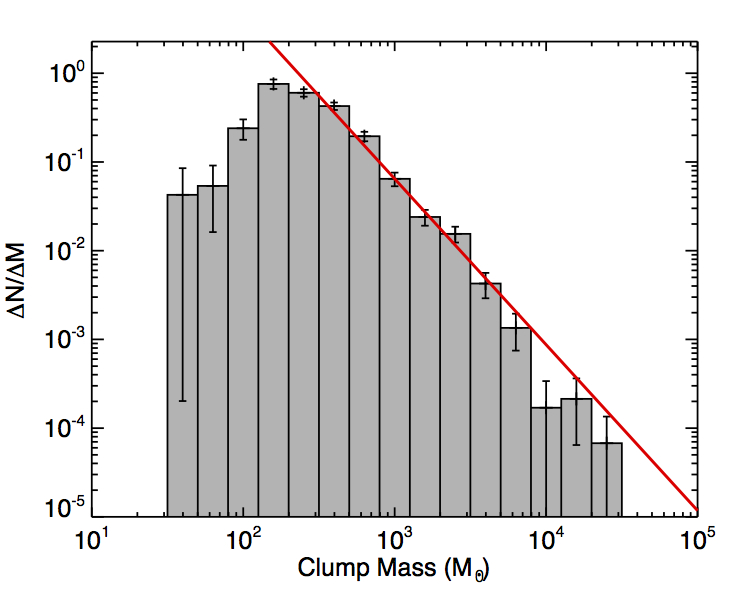}
\caption{The mass function of JPS compact sources associated with clouds in the W43 complex.  The red line indicates the least-squares linear fit to the data and the error bars indicate the Poisson statistics.}
\label{CMF}
\end{figure}

The W\,43 Giant Molecular Complex is the most distinctive feature in the JPS $\emph{l}$ = 30$\degr$ observations and dominates the emission within the region. This massive-star-forming complex is thought to be at the end of the Galactic Long Bar and molecular clouds associated with it cover longitudes of $\emph{l}$ = 29--32 and latitudes of $\mid$\,\emph{b}\,$\mid \le 1\degr$ \citep{nguyen11} and are found at velocities of 80--110 km\,s$^{-1}$.

By extracting spectra at the positions of the catalogued JPS sources from $^{13}$CO $J = 3 \rightarrow 2$ data cubes in the CO Heterodyne Inner Milky-Way Plane Survey (CHIMPS: Rigby et al., in preparation), which are briefly described in \citet{eden12}, or in $^{13}$CO $J=1 \rightarrow 0$ from the GRS, we can identify JPS sources that are in the $l,b,v$ range associated with the W\,43 complex. Because of the latitude range of the 
$^{13}$CO $J = 3 \rightarrow 2$ survey, GRS data were used for sources with $\mid$\,\emph{b}$\,\mid$ $>$ 0.5$\degr$.

676 JPS sources were found in the appropriate $\emph{l}$ and $\emph{b}$ range for W\,43, 47 of which required the extraction of GRS spectra to identify their velocities. 450 of these have velocities within the W\,43 limits and so are considered to be associated with the complex.

The masses of the sources were calculated using the standard optically thin conversion,

\begin{equation}
M = \frac{S_{\nu}D^{2}}{\kappa_{\nu}B_{\nu}(T_{d})}.
\end{equation}

\noindent
Taking the mass absorption coefficient at $\lambda$ =  850\,$\umu$m to be $\kappa_{850}$ = 0.001 m$^{2}$ kg$^{-1}$ \citep{mitchell01}, and assuming a distance of 5.49\,kpc (\citealp{zhang14}) and a single dust temperature of 18\,K for all sources, the conversion becomes $M/M_{\odot} = 225 \times S_{850}$/Jy.

We use the derived masses to plot the clump-mass function (CMF) of the W43 complex and this is presented in Figure\,\ref{CMF}. Assuming the CMF to be a power law above the observed turnover ($\log_{10}(M/M_{\odot})$ = 2.4) of the form $\Delta$$\emph{N/}$$\Delta$$\emph{M}$ $\propto$ $\emph{M}^{\alpha}$, a least-squares fit to the CMF gives an index of $\alpha = -1.87 \pm 0.05$. This is consistent with the results of \cite{urquhart14c}, who found values in the range $-1.85$ to $-1.91$ for ATLASGAL clumps that house tracers of massive-star formation, across different evolutionary stages and distributed over a large section of the Galactic Plane.

The slope of the distribution is also consistent with those found for Galactic giant molecular clouds (GMCs), although the mass range is very different. The spectrum of GMC masses greater than ($\log_{10}(M/M_{\odot})$ = 4) can also be fitted by a power law, with studies by \cite{solomon87}, \cite{heyer01} and \cite{roman10}, giving $-1.50\pm0.36$, $-1.80\pm0.03$ and $-2.02 \pm0.11$, respectively (the Roman-Duval et al.\ value has been re-calculated for $\Delta$$\emph{N/}$$\Delta$$\emph{M}$, rather than $\Delta$$\emph{N/}$$\Delta \log\emph{M}$, as published).

The fact that the mass functions of the clump and cloud populations are indistinguishable, both across the Galactic Plane and in a specific region as in this study, suggests that the same process, whether physical or stochastic, is determining both distributions. Measurements of the fraction of molecular gas within clouds that has been converted into dense clumps (the `clump-formation efficiency' or CFE), using the Bolocam Galactic Plane Survey (\citealt{eden12,eden13}; \citealt{battistiheyer}) found a fairly constant value between 5 and 15 per cent when averaged over samples of clouds.  This, together with a consistent power-law index for the mass functions of clouds and clumps, suggests that, outside of the Galactic Centre regions, the CFE is not sensitive to the local environment on kiloparsec scales (large variations in CFE from cloud to cloud are found on smaller scales, however; see \citealt{eden12}).

\subsection{Star Formation in the W43 sample}

\subsubsection{Virial State}

\begin{figure}
\includegraphics[width=0.47\textwidth, trim= 0 0 0 0]{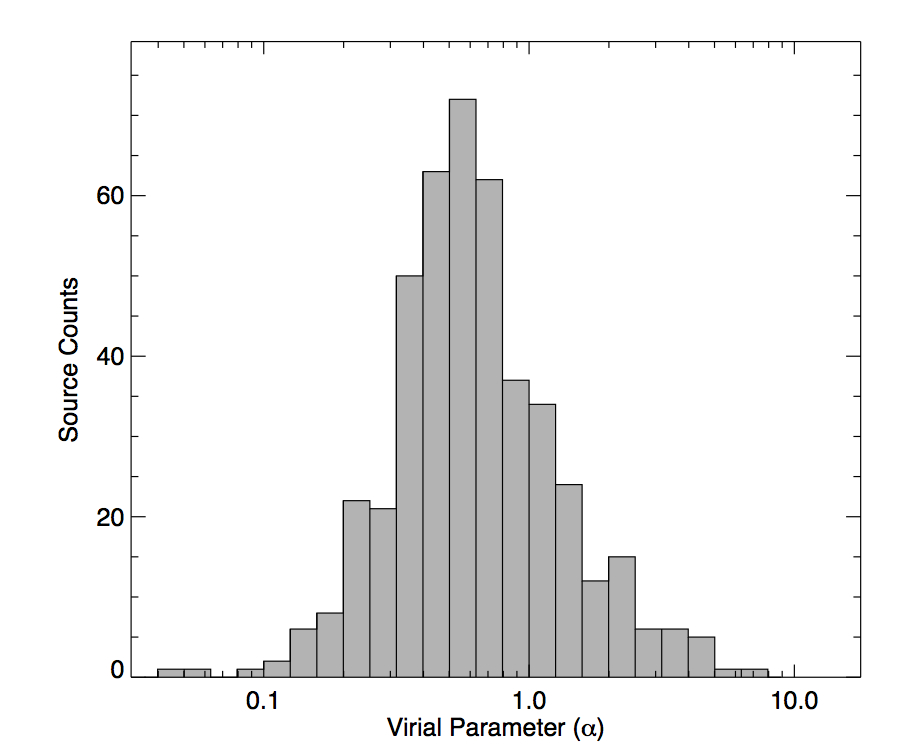}
\caption{Distribution of calculated virial ratios for the dense clumps identified with the W43 complex.}
\label{W43Mvirhist}
\end{figure}

\begin{figure}
\includegraphics[width=0.47\textwidth, trim= 0 0 0 0]{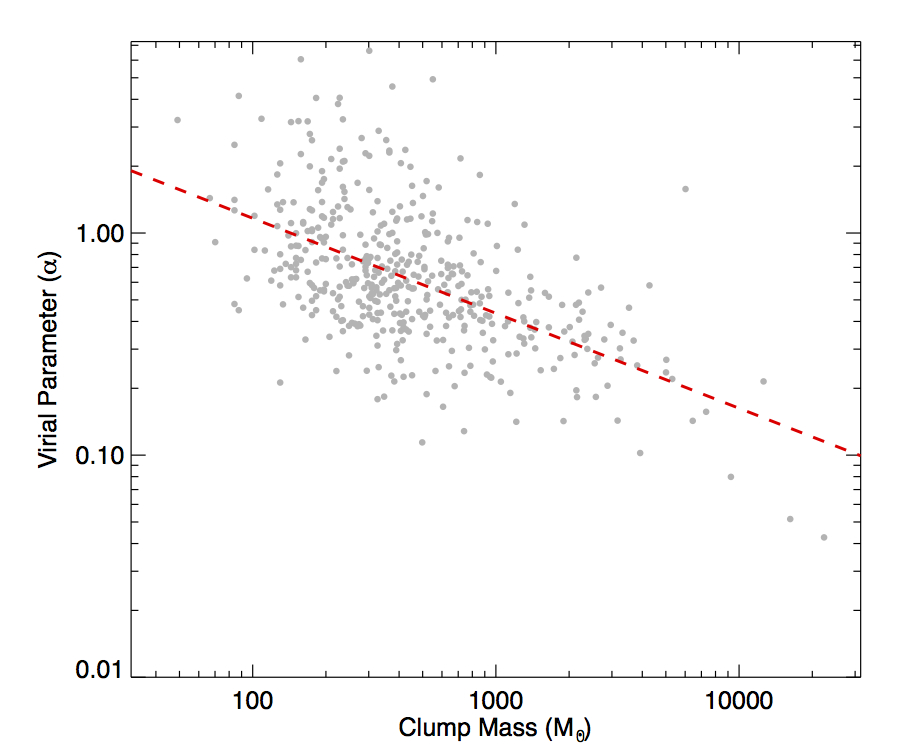}
\caption{Dependence of virial ratio on clump mass in the dense clumps identified with the W43 complex; The linear fit shown by the dashed line is $-0.42 \pm 0.08$ }
\label{W43Mvir_mass}
\end{figure}

To calculate the virial masses of the W43 sources, the measured $^{13}$CO line width was corrected to estimate the average velocity dispersion of the total column of gas.
Using the prescription of \cite{fullermyers}, the average line-of-sight velocity dispersion is

\begin{equation}
%\Delta v^{2}_{\rm avg} = 3 \Delta v^{2}_{\rm corr} + 8 \ln 2 \times \frac{kT_{\rm kin}}{m_{\rm H}}\left(\frac{1}{\umu_{\rm gas}} - \frac{1}{\umu_{\rm CO}}\right),
\Delta v^{2}_{\rm avg} = \Delta v^{2}_{\rm corr} + 8 \ln 2 \times \frac{kT_{\rm kin}}{m_{\rm H}}\left(\frac{1}{\umu_{\rm gas}} - \frac{1}{\umu_{\rm CO}}\right),
\label{vdispeqn}
\end{equation}

\noindent where $\Delta v^{2}_{\rm corr}$ is the observed CO line width corrected for the spectrometer resolution, $k$ is the Boltzmann constant, $T_{\rm kin}$ is the gas kinetic temperature, taken to be 12\,K from the average excitation temperatures found for W43 clouds from $^{13}$CO J=2--1 data by \citet{carlhoff13}.  The values in the latter work are derived from a single transition and are therefore lower limits.  The choice is not critical, however, as in any case the thermal line-width correction to the velocity dispersion in Equation \ref{vdispeqn} is small (no more than 10 per cent). 
The mean masses of molecular gas and of carbon monoxide are $\umu_{\rm gas} = 2.33$ and $\umu_{\rm CO} = 28$, respectively.  The virial mass for each clump can then be calculated from:

\begin{equation}
\left(\frac{M_{\rm vir}}{M_{\odot}}\right) = 161 \left(\frac{R}{\rm pc}\right)\left(\frac{\Delta v_{\rm avg}}{\rm km\,s^{-1}}\right)^{2},
\end{equation}

\noindent (\citealp{urquhart14c}) where $R$ is the half-power radius of the clump, evaluated from the geometric mean of $\sigma_{\rm maj}$ and $\sigma_{\rm min}$ from Table \ref{table:cattable}.

The distribution of the virial ratios ($\alpha$ = $M_{\rm vir}/M_{\rm clump}$ of the W43 clumps is shown  in Figure \ref{W43Mvirhist}. The distribution peaks just below $\alpha$ = 2 with a logarithmic mean of 
$-0.080$ and median of $-0.236$. 

It is a commonly observed result that virial ratios tend to cluster around a value within a factor of two of unity.  Allowing for scatter due to random error, this is usually taken to indicate that the majority of the clumps can be considered to be gravitationally bound and are likely to be either pre-stellar or protostellar.  It should be noted, however, that such results are probably subject to selection effects, especially that arising from the lack of stable equilibria between internal pressure and gravity, and the consequent short-lived nature of heavily gravitationally sub- or super-critical clump material that is observable in the 850-$\umu$m continuum, 

\cite{kauffmann13} (originally \citealp{bertoldi}) define the critical value for the virial ratio as $\alpha$ = 2 for sources which are not supported by a magnetic field and, as a result, sources with $\alpha < 2$ are gravitationally unstable and require the presence of a strong magnetic field to provide support. 
This is the case for 
93 per cent of the W43 sources. 

A strong relationship of decreasing $\alpha$ with increasing clump mass is found, similar to that seen in the results of \cite{kauffmann13} and by \cite{urquhart14c} for ATLASGAL clumps associated with tracers of massive star formation, and is shown in Figure \ref{W43Mvir_mass}. A fit to the data finds a slope of $-0.42 \pm 0.06$, consistent with \cite{urquhart14c}, who find $-0.53\pm0.16$.  This may again be the result of selection effects, particularly those associated with the mixing of gas and dust tracers in the analysis.  Taken at face value, however, this relationship suggests that the clumps in the sample are not quite Larson-like.  Such objects would have $\Delta v \propto M^{0.2}$ and $M \propto R^2$, which would give $\alpha \sim {\rm const}$. The observed relation requires either a steeper dependence of $M$ on $R$ or $\Delta v$ to be virtually constant, or both.

\subsubsection{Evolutionary State}

W43 is a site of recent and ongoing star formation. The scale of the current star formation activity can be determined by JPS clumps that are coincident with 70-$\umu$m Hi-GAL compact sources, the presence of which is taken to be a reliable indicator of star formation 
(e.g. Faimali et al. 2012).

A total of 287 70-$\umu$m sources were found to be coincident with 233 clumps from the W43 complex sample, with the majority of JPS clumps found to be coincident with a single Hi-GAL source (185 clumps). The largest number of Hi-GAL sources found to be associated with a single JPS clump was 3, which happened on 5 occasions.

The sample of massive clumps in W43 is complete above $\sim$100 M$_{\odot}$ and 440 of the whole sample of 450 have masses above this value.  If we assume that the star-formation rate is constant, the relative statistical lifetimes of the pre-stellar and star-forming clumps above this threshold is given by the relative numbers in the two subsets. By taking the lifetime of the clumps to be 0.5\,Myr (e.g.\ \citealp{ginsburg12}), we find that the relative lifetimes of the two phases are 0.26 $\times$ 10$^{6}$ yr and 0.24 $\times$ 10$^{6}$ yr for the star-forming and non-star-forming clumps, respectively.
It is no surprise that the lifetimes of the two phases are very similar as the {\em starless} lifetime of an eventual star-forming clump has been estimated at less than half a Myr (e.g.\ \citealp{ginsburg12}) and the timescale of the IR-bright stage of massive young stellar objects ranges from 7 $\times$ 10$^{4}$ yr to 4 $\times$ 10$^{5}$ yr, for sources with luminosities between 10$^{4}$ L$_{\odot}$ and 10$^{5}$ L$_{\odot}$ (\citealp{mottram11}), respectively. 

\begin{figure}
\includegraphics[width=0.47\textwidth, trim= 0 0 0 0]{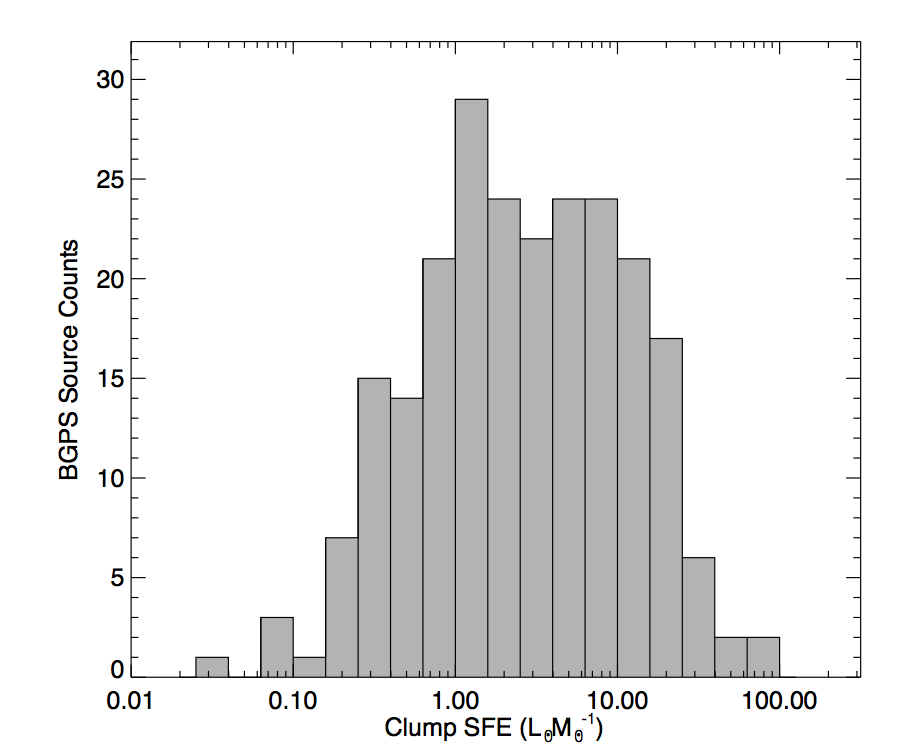}
\caption{The distribution of the ratio between clump (IR) luminosity to clump mass for the subset of sources in the W43 JPS sample that have associated 70-$\umu$m detections in the Hi-GAL data.}
\label{W43LMhist}
\end{figure}

In a sample-based study, the luminosity of the YSOs detected at 70\,$\umu$m can be usefully estimated from the mean ratio of the 70-$\umu$m flux to integrated infrared YSO luminosity obtained from the sample of \citet{eden15}. This ratio was 
measured to be $\log_{10}(F_{\rm bol}/F_{70\,\umu {\rm m}}) = 1.96$ with a standard deviation of 0.56.  The resulting luminosities can then be used to calculate the ratio $L/M$ for the individual clumps.  The resulting distribution of $L/M$ values is shown in Figure \ref{W43LMhist}. 

The range of values matches those found by \citet{eden15}, varying between 0 and 80 $L_{\odot}/M_{\odot}$, with a mean value of 6.50 $\pm$ 0.66 $L_{\odot}/M_{\odot}$ and a median value of 2.56 $\pm$ 2.06 $L_{\odot}/M_{\odot}$. The distribution of $L/M$ is consistent with being log-normal with both Anderson-Darling and Shapiro-Wilk tests giving probabilities of 0.107 and 0.151, respectively. The distribution can therefore be fitted with a Gaussian which has a mean of $\log_{10} (L/M) = 0.48$ (i.e.\ $L/M = 3.0$) and a standard deviation of 1.14, giving a standard error on the mean of 0.075 or about 20\%. These representative values are close to those found in an analysis of Hi-GAL data on infrared dark clouds in the longitude range $15\degr \le \ell \le 55\degr$ by \cite{traficante15}, who found mean $L/M$ values of $\sim$3.6 and $\sim$5.7 for starless and protostellar sources, respectively. The presence of a log-normal distribution tells us that the processes behind the distribution of $L/M$ values, both in this region and across the Galactic Plane, are the result of the multiplicative effect of multiple random processes within star-forming regions, be these the larger molecular-cloud environments (indicated by the log-normal distribution of clump-mass to cloud-mass ratios: \citealt{eden12,eden13}), or individual star-forming clumps.

\begin{figure}
\includegraphics[width=0.47\textwidth, trim= 0 0 0 0]{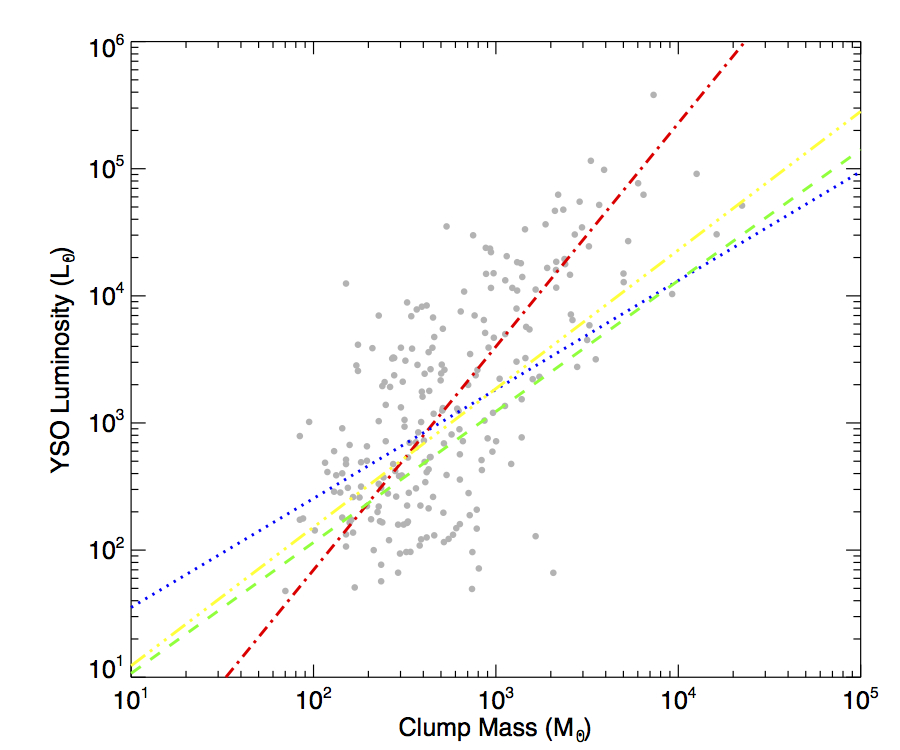}
\caption{
The relationship between embedded IR luminosity and clump mass for the subset of sources in the W43 JPS
sample that have associated 70-$\umu$m detections in the Hi-GAL data.  The red dot-dashed line indicates the linear fit to the JPS data.  Also shown are the relationships found by \citep{urquhart14c} for ATLASGAL sources associated with: any tracer of massive star formation (green dashed line) and a limited-distance subsample (yellow dot-dash); methanol masers only (blue dotted line). The gradients of those lines are 1.03, 1.09 and 0.857, respectively.
}
\label{W43LMplot}
\end{figure}

Figure \ref{W43LMplot} shows the relationship between $L$ and $M$ for the subset of clumps associated with the W43 cloud.  The gradient of the bisection-method linear regression fit
to the distribution of points in log space is $1.76 \pm 0.16$. This is 
significantly steeper than the value of $1.08\pm0.13$ found by \cite{urquhart14c}, for a distance-limited set of ATLASGAL clumps associated with tracers of massive star formation (methanol masers, H{\sc ii} regions and RMS MYSOs).  The difference may be accounted for by the inclusion of a population of sources with low $L/M$ values at masses below $\sim$$10^3$\,\Msun, that may be relatively unevolved.  

The mean value of $L/M$ in W43 is significantly lower than that of $16.0 \pm 0.7$ found for all ATLASGAL clumps associated with massive star formation but is closer to the $6.8\pm0.7$ found for methanol-maser-associated clumps alone by \cite{urquhart14c}.  $L/M$ is only weakly dependent on $M$, if at all, so the inclusion of higher-mass clumps in the ATLASGAL sample should not account for this.  The distribution in Figure \ref{W43LMplot} suggests rather that the W43 sample includes an excess population of sources in the mass range 100 to 1,000\,\Msun\ with low $L/M$ values.  These sources may be relatively young.  \citet{nguyen11} found that the `future' (i.e.\ incipient) SFR in W43, as traced by the sub-mm continuum, 
is around 5 times larger than the current SFR estimated from the 8-$\umu$m luminosity, and the larger value is consistent with the similar prediction by \citet[and see also \citealp{nguyen13}]{motte03}.

\section{summary}

Early results from the JCMT Plane Survey, part of the James Clerk Maxwell Telescope Legacy Survey programme are presented from the survey region centred on Galactic longitude $\ell = 30\degr$.  The results incorporate data at about the 40 per cent completion stage (in both this region and the whole survey), the point at which the sensitivity appears close to the confusion limit in this crowded area of the Galactic Plane.

The contamination of the 850-$\umu$m continuum band by emission in the $^{12}$CO J=3--2 line was investigated using COHRS data.  The majority of the significant continuum emission has contamination levels in the region of a few per cent, less than the flux calibration uncertainty, and insufficient to warrant the development of a detailed correction method at this time.

Compact-source identification and extraction are done using the \FW\ algorithm resulting in a sample of 1029 objects above a 5-$\sigma$ flux threshold.  The measured peak flux densities in this sample are compared to those in positionally matched detections from the ATLASGAL 870-$\umu$m compact-source catalogue and found to be consistent after correction for the different beam sizes.

Initial completeness tests were made using \FW\ to retrieve artificial sources from a uniform Gaussian noise field with rms deviation equivalent to that in the data.  The source flux densities were recovered accurately above peak values of 0.3 mJy per beam.  Below this, fluxes tended to be over-estimated, perhaps due to noise-related biasing in the detection method.  Source flux densities below 0.3 mJy\,beam$^{-1}$ have been corrected for this effect.  The completeness limit estimated in this way is of 95\% completeness above 80\,mJy\,beam$^{-1}$, which is around 4$\sigma$ in these data.  The true completeness limit in the $\ell = 30\degr$ region is larger than this, however, probably due to confusion.  We estimate this level to be about 0.2\,mJy\,beam$^{-1}$. The corresponding limit in the integrated flux densities is $\sim$0.8\,mJy which may be compared to a limit of 4--8\,Jy in the ATLASGAL survey.

The population of dense clumps associated with the W43 massive-star-forming complex clump masses is analysed.  The virial state of these clumps is clustered around the equilibrium value.  The relative statistical lifetime of the protostellar and starless clumps (with and without a 70-$\umu$m detection) are about equal and the distribution of the ratio $L/M_{\rm clump}$ suggests that these sources are younger than the Milky-Way average.

\section*{acknowledgements}

The James Clerk Maxwell Telescope has historically been operated by the Joint Astronomy Centre on behalf of the Science and Technology Facilities Council of the United Kingdom, the National Research Council of Canada and the Netherlands Organisation for Scientific Research.  The data presented in this paper were taken under JCMT observing programme MJLSJ02.

%\bibitem[\protect\citeauthoryear{author names} {year}]{key} Text of reference ...

%\bibitem[\protect\citeauthoryear{three author names} {first author etal}{year}]{key} Text of reference ... 

%\appendix

\clearpage
\onecolumn

\noindent
Author afilliations\\
$^{1}$Astrophysics Research Institute, Liverpool John Moores University, Ic2 Liverpool Science Park, 146 Brownlow Hill Liverpool \\ L3 5RF, UK\\
$^{2}$Department of Physics \& Astronomy, University of Calgary, 2500 University Drive NW, Calgary, AB, T2N1N4, Canada\\
$^{3}$Centre for Astrophysics Research, Science \& Technology Research Institute, University of Hertfordshire, College Lane, Hatfield, Herts, AL10 9AB, UK \\
$^{4}$Joint Astronomy Centre, 660 N. A'ohoku Place, University Park, Hilo, Hawaii 96720, USA\\
$^{5}$Max-Planck-Institut f\"ur Radioastronomie, Auf dem H\"ugel 69, 53121 Bonn, Germany \\
$^{6}$Observatoire astronomique de Strasbourg, Universit\'e de Strasbourg, CNRS, UMR 7550, France\\
$^{7}$Astrophysics Group, Cavendish Laboratory, J J Thomson Avenue, Cambridge, CB3 0HE, UK\\
$^{8}$Kavli Institute for Cosmology, Institute of Astronomy, University of Cambridge, Madingley Road, Cambridge, CB3 0HA, UK\\
$^{9}$Astrophysics Group, School of Physics, University of Exeter, Stocker Road, Exeter EX4 4QL, UK\\
$^{10}$Department of Physics and Astronomy, James Madison University, MSC 4502-901 Carrier Drive, Harrisonburg, VA 22807, USA\\
$^{11}$Department of Physics, University of Waterloo, Waterloo, Ontario N2L 3G1, Canada\\
$^{12}$School of Physics and Astronomy, E C Stoner Building, University of Leeds, Leeds LS2 9JT, UK\\
$^{13}$Leiden Observatory, Leiden University, P.O. Box 9513, 2300 RA Leiden, The Netherlands \\
$^{14}$School of Physics and Astronomy, Cardiff University, Cardiff CF24 3AA, UK\\
$^{15}$University of Athens, Department of Astrophysics, Astronomy and Mechanics, Faculty of Physics, Panepistimiopolis, \\ 15784 Zografos, Athens, Greece\\
$^{16}$501 Space Sciences, Cornell University, Ithaca, NY 14853 USA\\
$^{17}$Istituto di Astrofisica e Planetologia Spaziali (IAPS-INAF), via Fosso del Cavaliere 100, 00133, Roma, Italy\\
$^{18}$Jodrell Bank Centre for Astrophysics, School of Physics and Astronomy, The University of Manchester, Oxford Road, Manchester M13 9PL, UK\\
$^{19}$Centre de recherche en astrophysique du Qu\'ebec and D\'epartment de Physique, Universit\'e de Montr\'eal, Montr\'eal, H3C 3J7, Canada\\
$^{20}$NRC Herzberg Astronomy and Astrophysics, 5071 West Saanich Road, Victoria, BC, V9E 2E7, Canada\\
$^{21}$Astrophysics Group, Keele University, Keele, Staffordshire ST5 5BG, UK\\
$^{22}$Department of Physics \& Astronomy, University of British Columbia, 6224 Agricultural Road, Vancouver, BC V6T 1Z1, Canada\\
$^{23}$Department of Physics and Astronomy, Western Kentucky University, Bowling Green, KY 42101, USA\\
$^{24}$Joint ALMA Observatory, 3107 Alonso de Cordova, Vitacura, Santiago, Chile\\
$^{25}$Department of Physics and Astronomy, University of Victoria, Victoria, BC, V8P 1A1, Canada\\
$^{26}$D\'epartement de physique, de g\'enie physique et d'optique, Centre de Recherche en Astrophysique du Qu\'ebec, Universit\'e Laval, QC G1K 7P4, Canada\\
$^{27}$Canadian Institute for Theoretical Astrophysics, University of Toronto, 60 St. George Street, Toronto, ON M5S 3H8, Canada\\
$^{28}$Sydney Institute for Astronomy, School of Physics, The University of Sydney, NSW 2006, Australia\\
$^{29}$SRON Netherlands Institute for Space Research, University of Groningen PO-Box 800, 9700 AV Groningen The Netherlands\\
$^{30}$Kapteyn Astronomical Institute, PO Box 800, NL-9700 AV Groningen, the Netherlands\\
$^{31}$Department of Physics and Astronomy, University College London, WC1E 6BT London, UK\\
$^{32}$Universit\"at Bamberg, Markusplatz 3, Bamberg, 96045 Germany\\
$^{33}$Department of Physical Sciences, The Open University, Milton Keynes MK7 6AA, UK\\
$^{34}$RALSpace, The Rutherford Appleton Laboratory, Chilton, Didcot OX11 0NL, UK\\
$^{35}$National Astronomical Observatories, Chinese Academy of Science, 20A Datun Road, Chaoyang District, Beijing 100012, China\\

\label{lastpage}

\end{document}